\documentclass[twocolumn,superscriptaddress,showpacs,preprintnumbers,amsmath,amssymb]{revtex4}

\usepackage{booktabs}
\usepackage{graphicx,color}
\usepackage{dcolumn}
\usepackage{bm}
\usepackage{CJK}
\usepackage{mathrsfs}
\usepackage[pdfstartview=FitH,
            CJKbookmarks=true,
            bookmarksnumbered=true,
            bookmarksopen=true,
            colorlinks,
            pdfborder=001,
            linkcolor=blue,
            anchorcolor=blue,
            citecolor=blue
            ]{hyperref}
\usepackage{doi}

\usepackage{hyperref}
\usepackage[normalem]{ulem}

\begin{document}
\begin{CJK}{UTF8}{gbsn}

\title{Particle-vibration coupling for giant resonances beyond the diagonal approximation}

\author{Shihang Shen (申时行)}
\affiliation{Dipartimento di Fisica, Universit\`a degli Studi di Milano, Via Celoria 16, I-20133 Milano, Italy}
\affiliation{INFN, Sezione di Milano, Via Celoria 16, I-20133 Milano, Italy}

\author{Gianluca Col\`o \footnote{Email: Gianluca.Colo@mi.infn.it}}
\affiliation{Dipartimento di Fisica, Universit\`a degli Studi di Milano, Via Celoria 16, I-20133 Milano, Italy}
\affiliation{INFN, Sezione di Milano, Via Celoria 16, I-20133 Milano, Italy}

\author{Xavier Roca-Maza}
\affiliation{Dipartimento di Fisica, Universit\`a degli Studi di Milano, Via Celoria 16, I-20133 Milano, Italy}
\affiliation{INFN, Sezione di Milano, Via Celoria 16, I-20133 Milano, Italy}

\date{\today}

\begin{abstract}
A self-consistent particle-vibration coupling (PVC) model without diagonal approximation is presented.
The diagonal approximation, that neglects completely the interaction between the doorway states, has been removed by taking into account the interaction between the two particle-holes inside the doorway states.
As applications, isoscalar giant monopole, dipole, and quadrupole resonances in $^{16}$O are investigated based on the 
use of Skyrme functionals. The diagonal approximation is found to clearly impact on the strength distribution of the giant quadrupole resonance, and the description of the experimental data is improved without this approximation. The impact of the diagonal approximation is analyzed in detail, especially its effect on the eigenenergies and the induced coupling between neutron and proton particle-hole configurations. The latter is a direct and physically sound effect of the improvement on our formalism. The importance of using self-consistently the full effective interaction in the PVC vertex, and the effect of its renormalization via the subtraction method are also discussed. For completeness, we also analyze the dependence of our results on the Skyrme parameterization.\\
\end{abstract}

\maketitle

\section{Introduction}

Probing the response of a nucleus to the scattering of a particle or photon is a powerful tool to study the underlying nuclear structure. In the excitation energy range from 10 MeV to 30 MeV the nuclear systems show prominent and broad resonances, that are called giant resonances. Giant resonances have been experimentally studied for a long time \cite{Harakeh2001}, and yet the techniques that are developed 
are still improving towards unprecedented and advanced levels \cite{Gales2018,Bracco2019}. These studies provide extremely rich information
on the nuclear phenomenology. To name a few highlights, we mention the study of compression modes such as the Isoscalar Giant Monopole and Dipole Resonances that are undertaken in order to understand the incompressibility of uniform nuclear matter \cite{Garg2018}; the Isovector Giant Dipole Resonance and the associated dipole polarizability that is studied due to its implication for the symmetry energy \cite{Trippa2008,Roca-Maza2013,Roca-Maza2015,Bracco2019}; the low-lying dipole strength in the Isovector Dipole channel for its possible relation to the neutron skin thickness \cite{Wieland2011,Roca-Maza2012a,Savran2013,Bracco2015,Burrello2018}; the Isoscalar Giant Quadrupole Resonance which is tightly connected to the nucleon effective mass close to the Fermi surface \cite{Roca-Maza2013a}; the Gamow-Teller resonance \cite{Fujita2011} for its key role in astrophysically relevant weak-interaction processes \cite{Langanke2003} (cf. also the general discussion about giant resonances and the parameters of the nuclear equation of state in Ref.~\cite{Roca-Maza2018a}, and the report on giant resonances in nuclei far from stability 
in Ref.~\cite{Paar2007}).

On the other hand, the rich information on giant resonances also sets a challenge for theoretical descriptions. In the random phase approximation (RPA), the giant resonances are described as coherent superpositions of one particle one hole (1p-1h) excitations. The centroid of giant resonances and the energy weighted sum rule (EWSR) can be well described. However, the experimental resonance width $\Gamma$, which directly relates with the 
lifetime $\tau \equiv \frac{\hbar}{\Gamma}$, cannot be described by RPA due to the missing of the coupling with more complicated correlations.

Two main effects were identified to contribute to the width, the escape of a nucleon from the nucleus (escape width) and the spreading of the excitation energy into more complicated configurations (spreading width) \cite{Bertsch1983}.
Different efforts have been made to take into account these effects, for example, in second RPA (SRPA) the coupling of 1p-1h excitations with two particle two hole (2p-2h) excitations is taken into account \cite{Papakonstantinou2009,Papakonstantinou2010,Gambacurta2010,Gambacurta2018,Vasseur2018}.
In quasiparticle-phonon model, the excited states are composed of two-phonon excitations \cite{Soloviev1992,VanGiai1998,Iudice2012,Severyukhin2018}.
The quasiparticle-phonon model based on time-blocking approximation (TBA) \cite{Tselyaev2007,Lyutorovich2018a,Tselyaev2018} and its relativistic extension (RTBA) \cite{Litvinova2007,Litvinova2018,Robin2018} are developed within the many-body Green's function formalism, in which the 1p-1h (or two quasiparticles) $\otimes$ phonon configurations are included and more complicated intermediate states are blocked.
In the equation-of-motion phonon method, a set of equations for multiphonon states are derived, within the Tamm-Dancoff approximation (TDA).

Recently, a systematic approach to the response functions with nonperturbative treatment of higher configurations is formulated in the equation-of-motion framework, with a truncation at the level of two-fermion correlation functions.
The PVC-TBA method is then compared here with the formulas of the equation-of-motion approach and this provides a guidance to develop a systematic treatment of the response functions \cite{Litvinova2018a}.

In this work we use the particle-vibration coupling (PVC) model, which takes into account the coupling between a nucleon and the low-lying nuclear collective excitations (phonons) \cite{Bohr1998b}. In early applications, phenomenological inputs were used for the PVC vertex and parameters were adjusted to reproduce the data, making it difficult to have a universal description \cite{Bortignon1977,Bertsch1983,Mahaux1985}. A self-consistent treatment for the interaction in the PVC vertex, on top of a mean field associated with Skyrme functionals was worked out in Ref.~\cite{Bernard1980}, although only the velocity-independent central term was included in the vertex. The approach was further developed in Refs.~\cite{Colo1992,Colo1994,Colo2010,CaoLG2014}, and now the full Skyrme interaction is used for both the PVC vertex and the mean field. The same consistency has been achieved also in the relativistic PVC \cite{Litvinova2006}.
The PVC has been extended to describe open-shell nuclei within the Hartree-Fock (HF) plus BCS framework \cite{Colo2001}, and later on in the Hartree-Fock-Bogoliubov framework \cite{NiuYF2016}. By including both collective and noncollective excitations, the so-called hybrid configuration mixing (HCM) model was developed to study the low-lying spectroscopy of odd nuclei, and shell-model-type states like 2p-1h can be well taken into account \cite{Colo2017}. To better understand the renormalization of the effective interaction, the subtraction method developed in Ref.~\cite{Tselyaev2007} has been studied in the PVC \cite{Roca-Maza2017}. The PVC has been used to investigate, for example, the $\beta$-decay \cite{NiuYF2015,NiuYF2018} and good descriptions were achieved.

However, in the above PVC studies the so-called diagonal approximation has been used, that is, the 1p-1h state coupled with a phonon, which is also called a doorway state, has no interaction with other doorway states. This is similar to the diagonal approximation in the SRPA where there is no interaction among the 2p-2h states.
In the context of SRPA, this approximation has been tested against the fully self-consistent framework, and it has been shown to affect significantly the strength distributions \cite{Gambacurta2010}.
Such correlations have also been considered in the TBA \cite{Tselyaev2007} and studied by RTBA \cite{Litvinova2010,Litvinova2013}, based on the equation-of-motion method \cite{Schuck1976}.
It was found that the two-phonon correlations push large part of the pygmy strength above the neutron threshold, in better agreement with available data for the tin isotopes and $^{68}$Ni \cite{Litvinova2010,Litvinova2013}.
Further progresses on higher-order correlations beyond the 2p-2h level of configuration can be found in Refs~\cite{Tselyaev2018,Litvinova2015}.

Therefore, it is of importance to have a closer view into the diagonal approximation in the PVC.
Here we adopt a different approach from the works by Refs. \cite{Litvinova2010,Litvinova2013}.
We will use the equation-of-motion method similar to the one used in the SRPA described in Ref.~\cite{Yannouleas1987}, and the two particle-holes inside the doorway states will interact through the particle-hole interaction.
We will compare the effect of removing diagonal approximation with the ones in SRPA \cite{Gambacurta2010} and RTBA \cite{Litvinova2010,Litvinova2013}.
An analytical comparison to the formalisms of RTBA will also be given in the Appendix.
The sum rules in current PVC framework will also be discussed both analytically and numerically.

In Sec. \ref{sec:theory}, we give a brief summary of the formalisms of the HF, RPA, and PVC. The numerical details for the calculations are discussed in Sec. \ref{sec:nd}. Results for the isoscalar giant monopole, dipole, and quadrupole resonances of $^{16}$O by PVC without diagonal approximation are presented in Sec. \ref{sec:res}. Finally, the summary and perspectives for future investigations will be given in Sec. \ref{sec:sum}.


\section{Theoretical Framework}\label{sec:theory}

\subsection{From Hartree-Fock to Random Phase Approximation}

Our starting point is the Skyrme functional which is constructed from the Skyrme effective interaction solved within the Hartree-Fock (HF) approximation. The detail of the Skyrme interaction and the corresponding formulas of the Skyrme Hartree-Fock theory in spherical nuclei have been given in detail \cite{Vautherin1972} and will not be repeated here. In this work we take the doubly magic nucleus $^{16}$O as an example, so that effects of pairing and deformation \cite{Vautherin1973} can be ignored. The Hartree-Fock ground state $|\Phi_0^{\rm HF}\rangle$ is a single Slater determinant.
In the second quantized form it can be written as:
\begin{equation}\label{eq:}
  |\Phi_0^{\rm HF}\rangle = \prod_i^A a_i^\dagger |\rangle,
\end{equation}
where $A$ is the number of nucleons of a given nucleus, $a_i^\dagger$ is the creation operator of HF single-particle state $|i\rangle$, and $|\rangle$ is the bare vacuum. The HF equation is solved with a box boundary condition and a set of discrete occupied and unoccupied states $|i\rangle$ are obtained. The Hamiltonian of the system can be expressed as
\begin{equation}\label{eq:}
  H = H_0 + V_{\rm res},
\end{equation}
where $H_0$ is the HF Hamiltonian and $V_{\rm res}$ the residual interaction:
\begin{align}
  H_0 &= \sum_i^A e_i a_i^\dagger a_i^{} - \frac{1}{2} \sum_{ij}^A \bar{V}_{ijij}, \\
  V_{\rm res} &= \frac{1}{4} \sum_{k'l'kl} \bar{V}_{k'l'kl} :a_{k'}^\dagger a_{l'}^\dagger a_l^{} a_k^{}:.
\label{eq:}
\end{align}
In the above equations, $e_i$ is the single-particle energy of state $|i\rangle$, and $\bar{V}_{ijij} = V_{ijij} - V_{ijji}$ is the antisymmetrized two-body matrix element. The normal ordered product of operators $a_{k'}^\dagger a_{l'}^\dagger a_l^{} a_k^{}$ is labelled as $:a_{k'}^\dagger a_{l'}^\dagger a_l^{} a_k^{}:$ with respect to the HF particle-hole vacuum $|\Phi_0^{\rm HF}\rangle$.

To study the excited state properties, one can use the RPA, in which all the possible 1p-1h excitations are considered. If we define the HF ground state $|\Phi_0^{\rm HF}\rangle$ and all the 1p-1h excitations $|ph\rangle$ built upon as the subspace $Q_1$, the RPA solution can be obtained by diagonalizing the Hamiltonian in this subspace $Q_1HQ_1$. For the derivation of the RPA equations and their solution we refer the reader to Ref. \cite{Ring1980}.
The RPA equation reads
\begin{equation}\label{eq:rpa}
  \sum_{ph} \left(\begin{array}{cc}
  A & B \\ -B^* & -A^*
  \end{array}\right)_{p'h',ph}
  \left(\begin{array}{c}
  X_{ph}^{(n)} \\ Y_{ph}^{(n)}
  \end{array}\right)
  = \omega_n
  \left(\begin{array}{c}
  X_{p'h'}^{(n)} \\ Y_{p'h'}^{(n)}
  \end{array}\right)
\end{equation}
with $\omega_n$ the excitation energy of RPA state $|\Phi_n^{\rm RPA}\rangle$ (that can be simply labeled as $|n\rangle$ when there is no ambiguity), $X_{ph}^{(n)}$ and $Y_{ph}^{(n)}$ the corresponding RPA wave function coefficients. The matrix elements $A$ and $B$ are
\begin{subequations}\label{eq:ab}\begin{align}
  A_{p'h',ph} &= \langle 0|[a_{h'}^\dagger a_{p'}^{}, [H, a_{p}^\dagger a_{h}^{} ]] |0 \rangle, \notag \\
  &= \delta_{p'h',ph} (e_p - e_h) + \bar{V}_{p'hh'p}, \\
  B_{p'h',ph} &= -\langle 0|[a_{h'}^\dagger a_{p'}^{}, [H, a_{h}^\dagger a_{p}^{} ]] |0 \rangle = \bar{V}_{p'ph'h},
\end{align}\end{subequations}
where $|0\rangle$ is the RPA ground state $|\Phi_0^{\rm RPA}\rangle$, and within the quasiboson approximation it is replaced by the HF ground state $|\Phi_0^{\rm HF}\rangle$ \cite{Ring1980}. Without causing confusion, the simple form $|0\rangle$ of the ground state will be used later on also in the framework of PVC. The RPA excited states, or the phonons, can be expressed as
\begin{equation}\label{eq:}
  |n\rangle = Q_n^\dagger |0\rangle,
\end{equation}
with
\begin{equation}\label{eq:qn}
  Q_n^\dagger = \sum_{ph} \left[ X_{ph}^{(n)} a_p^\dagger a_h^{} - Y_{ph}^{(n)} a_h^\dagger a_p^{} \right],
\end{equation}
and the RPA ground state satisfies
\begin{equation}\label{eq:}
  Q_n |0\rangle \equiv 0.
\end{equation}

\subsection{Particle-vibration coupling}

As we briefly mentioned in the Introduction,
RPA can give a good description of the centroid energy of giant resonances as well as of the EWSR exhausted by each mode.
However, properties such as the width of the resonances cannot be well described. Part of the width comes from the so called Landau damping effect and part of it is due to correlations beyond 1p-1h \cite{Bertsch1983}. The Landau damping effect 
produces a fragmentation of the strength, in contrast with the ideal situation in which there is a single collective peak. Such an effect depends on the intensity of the residual interaction that 1p-1h configurations feel, as well as the density of the unperturbed 1p-1h states around the 
resonance energy.
Coupling with more complicated states than 1p-1h produce the resonance spreading width.
Our formalism can also account for the other mechanism giving rise to the resonance width, since the escape of a nucleon can be also described. 

To take into account these effects, two subspaces $P$ and $Q_2$ are built. Similar to $Q_1$, subspace $P$ is made up with 1p-1h configurations but now the particle is a continuum state and orthogonal to all the states in $|i\rangle$. For subspace $Q_2$, one can chose the 2p-2h configurations and the resulting framework would be the second RPA \cite{Yannouleas1983}. In the particle-vibration coupling model, the $Q_2$ space is composed of the so-called doorway states $|N\rangle$ with 1p-1h excitation coupled to a RPA phonon,
\begin{equation}\label{eq:doorway}
  |N\rangle = |ph\rangle \otimes |n\rangle.
\end{equation}
The corresponding excitation operator reads
\begin{equation}\label{eq:QN}
  \tilde{Q}_N^\dagger = \sum_{ph,n} \left[ \tilde{X}_{ph,n}^{N} a_p^\dagger a_h^{} Q_n^\dagger - \tilde{Y}_{ph,n}^{N} Q_n^{} a_h^\dagger a_p^{} \right].
\end{equation}

Now, the PVC equation is an eigenequation in the $P+Q_1+Q_2$ space,
\begin{equation}\label{eq:}
  H (P+Q_1+Q_2)\Psi = \omega (P+Q_1+Q_2)\Psi,
\end{equation}
$\Psi$ being the full-space wave function to be projected out.
After truncating higher orders, this equation can be mapped into $Q_1$ with an energy dependent Hamiltonian as \cite{Colo1994} (see Appendix \ref{app:hq1})
\begin{equation}\label{eq:pvceq}
  \mathcal{H}(\omega) Q_1\Psi = \left( \Omega_\nu - i\frac{\Gamma_\nu}{2} \right) Q_1\Psi.
\end{equation}
Both the effective Hamiltonian $\mathcal{H}$ and the eigensolutions are complex.
The effective Hamiltonian is composed of three terms,
\begin{widetext}
\begin{equation}\label{eq:homega}
  \mathcal{H}(\omega) \equiv Q_1HQ_1 + W^\uparrow(\omega) + W^\downarrow(\omega)
  = Q_1HQ_1 + Q_1HP\frac{1}{\omega-PHP+i\epsilon} PHQ_1 + Q_1HQ_2\frac{1}{\omega-Q_2HQ_2+i\epsilon} Q_2HQ_1,
\end{equation}
\end{widetext}
i.e., the RPA term, escape term ($W^\uparrow$), and spreading term ($W^\downarrow$).
For the calculation of the escape term, one is referred to Ref. \cite{Colo1994}.
For more detail of the spreading term and the diagonal approximation of it, see Section \ref{sec:spread}.

As one is now working in the $Q_1$ subspace, the RPA solutions can be used as a basis to expand the PVC state as
\begin{equation}\label{eq:wf}
  |\nu\rangle = \sum_n F_n^{(\nu)} |n\rangle.
\end{equation}
Then the PVC equation (\ref{eq:pvceq}) takes the matrix form
\begin{equation}\label{eq:pvceq2}
  \sum_n \mathcal{H}_{n'n}(\omega) F_{n}^{(\nu)} = \left( \Omega_\nu - i\frac{\Gamma_\nu}{2} \right) F_{n'}^{(\nu)},
\end{equation}
with
\begin{equation}\label{eq:hnn}
  \mathcal{H}_{n'n}(\omega) = \omega_n + W_{n'n}^\uparrow(\omega) + W_{n'n}^\downarrow(\omega).
\end{equation}
The matrix of the wave function coefficients is complex orthogonal,
\begin{equation}\label{eq:}
  F^TF = FF^T = 1.
\end{equation}

The polarizability associated with the operator $O$ is defined as
\begin{equation}\label{eq:polar}
  \Pi(\omega) = \langle 0| O^\dagger \frac{1}{\omega-\mathcal{H}(\omega)+i\epsilon}O|0 \rangle.
\end{equation}
The corresponding strength function is
\begin{align}
  S(\omega) &= -\frac{1}{\pi} {\rm Im} \Pi(\omega) \notag \\
  &= -\frac{1}{\pi} {\rm Im}\sum_\nu \langle 0|O|\nu \rangle^2 \frac{1}{\omega-\Omega_\nu+i\frac{\Gamma_\nu}{2}}.
\label{eq:str}
\end{align}
The sum rules, or the $k$th moments $m_k$ of the strength function, are defined as
\begin{equation}\label{eq:mk}
  m_k = \int_0^\infty S(\omega) \omega^k d\omega.
\end{equation}
Among them, the energy-weighted sum rule $m_1$ is of particular interest as it can be expressed in a simple form via a double commutator evaluated in the ground state, namely
\begin{equation}\label{eq:dbc}
  m_1 = \frac{1}{2} \langle 0|[O^\dagger,[H,O]]|0 \rangle.
\end{equation}

\subsection{Spreading term in PVC}\label{sec:spread}

The spreading term is the last term in Eq.~(\ref{eq:homega}),
\begin{equation}\label{eq:wd}
  W^\downarrow(\omega) = Q_1HQ_2\frac{1}{\omega-Q_2HQ_2+i\epsilon} Q_2HQ_1.
\end{equation}
It describes the process in which 1p-1h configurations of the $Q_1$ subspace are coupled to the more complicated doorway states of the $Q_2$ subspace.
These terms can be derived with the equation-of-motion method \cite{Rowe1968} as in the SRPA~\cite{Yannouleas1987}.
Similar to the RPA matrix $Q_1HQ_1$ in Eq.~(\ref{eq:rpa}), one has the matrix $Q_1HQ_2$ and $Q_2HQ_2$ in the particle-hole and phonon representation:
\begin{align}
  Q_1HQ_2 &=
  \left(\begin{array}{cc}
  A_{ph,p_1h_1n} & B_{ph,p_1h_1n} \\
  -B_{ph,p_1h_1n}^* & -A_{ph,p_1h_1n}^* \\
  \end{array}\right) \label{eq:q1hq2} \\
  Q_2HQ_2 &=
  \left(\begin{array}{cc}
  A_{p_1h_1n_1,p_2h_2n_2} & B_{p_1h_1n_1,p_2h_2n_2} \\
  -B_{p_1h_1n_1,p_2h_2n_2}^* & -A_{p_1h_1n_1,p_2h_2n_2}^* \\
  \end{array}\right) 
\label{eq:q2hq2}
\end{align}
with the matrix elements defined similarly to Eq.~(\ref{eq:ab}),
\begin{align}
  A_{ph,p_1h_1n} &= \langle 0|[a_h^\dagger a_p^{}, [H, a_{p_1}^\dagger a_{h_1}^{} Q_n^\dagger ]] |0 \rangle, \\
  B_{ph,p_1h_1n} &= -\langle 0|[a_h^\dagger a_p^{}, [H, Q_n^{} a_{h_1}^\dagger a_{p_1}^{} ]] |0 \rangle, \\
  A_{p_1h_1n_1,p_2h_2n_2} &= \langle 0|[Q_{n_1}^{} a_{h_1}^\dagger a_{p_1}^{}, [H, a_{p_2}^\dagger a_{h_2}^{} Q_{n_2}^\dagger ]] |0 \rangle, \\
  B_{p_1h_1n_1,p_2h_2n_2} &= -\langle 0|[Q_{n_1}^{} a_{h_1}^\dagger a_{p_1}^{}, [H, Q_{n_2}^{} a_{h_2}^\dagger a_{p_2}^{} ]] |0 \rangle.
\label{eq:b22}
\end{align}
They can be evaluated as
\begin{align}
  A_{ph,p_1h_1n} &= \delta_{hh_1} \langle p|V|p_1,n \rangle - \delta_{pp_1} \langle h_1|V|h,n \rangle, \label{eq:a12} \\
  A_{p_1h_1n_1,p_2h_2n_2} &= \delta_{n_1n_2} \left[ \delta_{p_1h_1,p_2h_2} \left( \omega_{n_1} + e_{p_1h_1} \right) + \bar{V}_{p_1h_2h_1p_2} \right], \label{eq:a22} \\
  B_{ph,p_1h_1n} &= B_{p_1h_1n_1,p_2h_2n_2} = 0, \label{eq:b12}
\end{align}
with $\omega_n$ the energy of the phonon $|n\rangle$, $e_{ph} = e_p - e_h$, and
\begin{equation}\label{eq:vabn}
  \langle a|V|b,n \rangle = \sum_{ph} \left[ X_{ph}^{(n)} \bar{V}_{ahbp} + Y_{ph}^{(n)} \bar{V}_{apbh} \right].
\end{equation}

\begin{figure}[h]
  \includegraphics[width=8cm]{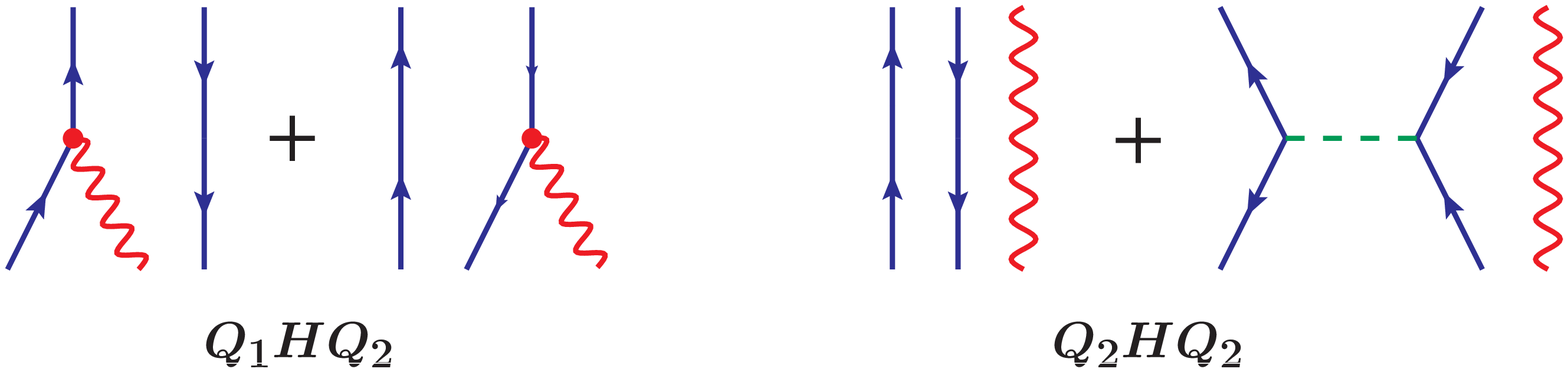}
  \caption{Schematic picture of the interaction of $Q_1HQ_2$ and $Q_2HQ_2$, corresponding to the matrix elements in Eqs.~(\ref{eq:a12}-\ref{eq:a22}). The straight lines are used to represent fermions (with up-arrow for a particle and down-arrow for a hole), while wave lines are for phonon states. The solid circle between two particle (or two hole) lines and a phonon is for the phonon vertex $\langle p|V|p1,n \rangle$ (or $\langle h1|V|h,n \rangle$) in Eq.~(\ref{eq:a12}), the dashed line is for the interaction between two particle-holes $\bar{V}_{p_1h_2h_1p_2}$ in Eq.~(\ref{eq:a22}).}
  \label{fig:q1hq2}
\end{figure}

The matrix element $A_{ph,p_1h_1n}$ in Eq.~(\ref{eq:a12}) represents the interaction between the 1p-1h state $|ph\rangle$ in the $Q_1$ space and the doorway state $|p_1h_1\rangle\otimes|n\rangle$ in the $Q_2$ space.
A diagrammatic representation of this interaction is given in the left part of Fig.~\ref{fig:q1hq2}, where straight lines are denoted for fermions (with up-arrow a particle and down-arrow a hole), red wave lines are for phonons. The solid circle between two particle (or two hole) lines and a phonon is for the phonon vertex $\langle p|V|p1,n \rangle$ (or $\langle h1|V|h,n \rangle$) in Eq.~(\ref{eq:a12}).
The matrix element $A_{p_1h_1n_1,p_2h_2n_2}$ in Eq.~(\ref{eq:a22}) represents the interaction among the doorway states, 
and its diagrammatic representation is also provided in the right part of Fig.~\ref{fig:q1hq2}.
The non-interacting part (first one) is denoted as $\delta_{n_1n_2} \delta_{p_1h_1,p_2h_2} \left( \omega_{n_1} + e_{p_1h_1} \right)$, and the dashed line in the interacting part (second one) is for the interaction between two particle-holes $\bar{V}_{p_1h_2h_1p_2}$ in Eq.~(\ref{eq:a22}).

The full spreading term can then be written as
\begin{align}
  W_{p'h',ph}^\downarrow(\omega) &= \sum_{p_1'h_1'p_1h_1n} A_{p'h',p_1'h_1'n} \notag \\
  &\times \left( \omega-A_{p_1'h_1'n,p_1h_1n} +i\epsilon\right)^{-1}
  A_{p_1h_1n,ph}.
\label{eq:wph}
\end{align}
In the above notation $\left( \omega-A_{p_1'h_1'n,p_1h_1n} +i\epsilon\right)^{-1}$ is not the inverse of a single matrix element, but the matrix element of the inverted matrix of $\omega-A_{p_1'h_1'n,p_1h_1n} +i\epsilon$. 

In previous investigations, the diagonal approximation was used, that is, no interaction among the doorway states was considered \cite{Colo1994}.
Within this approximation, the matrix element $A_{p_1h_1n_1,p_2h_2n_2}$ in Eq.~(\ref{eq:a22}) becomes
\begin{equation}\label{eq:a22dia}
  A_{p_1h_1n_1,p_2h_2n_2} = \delta_{n_1n_2} \delta_{p_1h_1,p_2h_2} \left( \omega_{n_1} + e_{p_1h_1} \right).
\end{equation}
The matrix $Q_2HQ_2$ then becomes diagonal, and the spreading term can be easily evaluated as
\begin{equation}\label{eq:wph2}
  W_{p'h',ph}^\downarrow(\omega) = \sum_{p_1h_1n} \frac{A_{ph,p_1h_1n}A_{p_1h_1n,ph}}{\omega-\omega_n-e_{p_1h_1}+i\epsilon}.
\end{equation}

When the diagonal approximation is not considered, there is an extra step of inverting the matrix $\omega - Q_2HQ_2 + i\epsilon$ before evaluating the spreading term.
See Appendix \ref{app:me} for more details.

Finally, interactions that are fitted at the mean-field level and are used within effective 
theories that go beyond mean field should in principle be refitted against to experimental data in order to avoid double-counting. That is, a renormalization of the model parameters is compulsory. The parameters will change their value since many-body contributions beyond mean-field are now explicitly included. The purpose of the subtraction method \cite{Tselyaev2007,Roca-Maza2017} is to provide a recipe for the renormalization of the effective interaction within the adopted model scheme that avoids a refitting of the parameters. For that, the spreading term in Eq.~(\ref{eq:homega}) should be replaced by
\begin{equation}\label{eq:sub}
  W^\downarrow(\omega) \to W^\downarrow(\omega) - W^\downarrow(\omega = 0).
\end{equation}

\subsection{Sum rules}\label{sec:sr}

In this subsection we discuss the sum rules.
Following a similar derivation from the response theory in the extended RPA \cite{Adachi1988}, the sum rules (\ref{eq:mk}) can be obtained as
\begin{equation}\label{eq:}
  m_k = \frac{1}{2} O^\dagger \mathscr{T} (\mathscr{M}^{-1} \mathscr{I})^k \mathscr{M}^{-1} \mathscr{T}^\dagger O,
\end{equation}
with $O$ the one-body excitation operator same as in Eq.~(\ref{eq:polar}),
\begin{equation}\label{eq:oij}
  O = \sum_{ij} O_{ij} a_i^\dagger a_j.
\end{equation}
$\mathscr{I}$ is the stability matrix, and $\mathscr{T}$ and $\mathscr{M}$ are the metric matrices,
\begin{align}
  \mathscr{T}_{c\beta} &=
  \left(\begin{array}{cc}
  \tilde{U}_{c\beta} & \tilde{V}_{c\beta}
  \end{array}\right), \quad
  \mathscr{M}_{\alpha\beta} =
  \left(\begin{array}{cc}
  {U}_{\alpha\beta} & {V}_{\alpha\beta} \\
  -{V}_{\alpha\beta} & -{U}_{\alpha\beta}
  \end{array}\right).
\label{eq:metric}
\end{align}
For $Q_1$ subspaces only, considering the case of $p'h'$ ($h'p'$ will be similar) one has
\begin{align}
  \tilde{U}_{c\beta} &: \tilde{U}_{p'h',ph} = \langle 0|[a_{h'}^\dagger a_{p'}, a_{p}^\dagger a_h]|0 \rangle = \delta_{p'h',ph}, \\
  \tilde{V}_{c\beta} &: \tilde{V}_{p'h',ph} = \langle 0|[a_{h'}^\dagger a_{p'}, a_{h}^\dagger a_p]|0 \rangle = 0, \\
  {U}_{\alpha\beta} &: {U}_{p'h',ph} = \langle 0|[a_{h'}^\dagger a_{p'}, a_{p}^\dagger a_h]|0 \rangle = \tilde{U}_{p'h',ph}, \\
  {V}_{\alpha\beta} &: {V}_{p'h',ph} = \langle 0|[a_{h'}^\dagger a_{p'}, a_{h}^\dagger a_p]|0 \rangle = \tilde{V}_{p'h',ph}.
\label{eq:}
\end{align}
The index $c$ is used to denote the pair $a_i^\dagger a_j$ in Eq. (\ref{eq:oij}).
The indices $\alpha$ and $\beta$ for $Q_1$ subspace are for $a_p^\dagger a_h$ or $a_h^\dagger a_p$ in Eq.~(\ref{eq:qn}); for $Q_2$ subspace they are for $a_p^\dagger a_h Q_n^\dagger$ and $Q_na_h^\dagger a_p$ in Eq.~(\ref{eq:QN}); for $P$ subspace it is similar to $Q_1$ but with the particle in the continuum, which we will label as
\begin{equation}\label{eq:qtilden}
  Q_{\tilde{n}}^\dagger = \sum_{\tilde{p}h} \left[ X_{\tilde{p}h}^{(\tilde{n})} a_{\tilde{p}}^\dagger a_h - Y_{\tilde{p}h}^{(\tilde{n})} a_h^\dagger a_{\tilde{p}} \right].
\end{equation}
When $P$ and $Q_2$ subspaces are included, the dimension of metric matrices in Eq.~(\ref{eq:metric}) will be enlarged accordingly.
It has been shown for SRPA \cite{Adachi1988} that even the 2p-2h correlations are considered in the excitation state, only the 1p1h components of $(\mathscr{M}^{-1} \mathscr{I})^k$ contribute to the sum rules, because of the absence of ground-state correlations.
This can be seen when one tries to evaluate the metric matrix elements with the ground state $|0\rangle$ chosen as the HF state $|\Phi_0^{\rm HF}\rangle$,
\begin{equation}\label{eq:}
  \tilde{U}_{ij,p_1p_2h_1h_2} = \langle 0|[a_{j}^\dagger a_{i}, a_{p_1}^\dagger a_{p_2}^\dagger a_{h_2}a_{h_1}]|0 \rangle = 0.
\end{equation}
In the end one can prove that the $m_0$ and $m_1$ are the same for SRPA and RPA \cite{Adachi1988}.
For extended RPA, however, the 2p-2h correlations are also included in the ground state $|0\rangle$ and in this case $\tilde{U}_{ij,p_1p_2h_1h_2}$ has non-zero components, pp ($\tilde{U}_{pp,p_1p_2h_1h_2}$) and hh ($\tilde{U}_{hh,p_1p_2h_1h_2}$).
As a result, $m_0$ and $m_1$ are different from RPA \cite{Adachi1988}.

In our PVC framework, the $Q_2$ subspace (\ref{eq:doorway},\ref{eq:QN}) is similar to the 2p-2h subspace in SRPA, and the ground state $|0\rangle$ is also chosen as the HF ground state.
It is then not difficult to find a similar conclusion as in SRPA,
\begin{equation}\label{eq:}
  \tilde{U}_{ij,phn} = \langle 0|[a_{j}^\dagger a_{i}, a_{p}^\dagger a_{h} Q_{n}^\dagger]|0 \rangle = 0,
\end{equation}
that is, the one-body excitation operator $O$ (\ref{eq:oij}) cannot connect the ground state to the $Q_2$ subspace in our framework.
Therefore, as a result, the $m_0$ and $m_1$ should be the same as in RPA.
This also agrees with the TBA, that when the effective interaction coincides with the one of RPA, the EWSR is the same as RPA \cite{Tselyaev2007}.

For the $P$ subspace (\ref{eq:qtilden}), the following term in the metric matrix is non-zero
\begin{equation}\label{eq:}
  \tilde{U}_{\tilde{p}'h',\tilde{p}h} = \langle 0|[a_{h'}^\dagger a_{\tilde{p}'}, a_{\tilde{p}}^\dagger a_{h}]|0 \rangle = \delta_{\tilde{p}'h',\tilde{p}h}.
\end{equation}
While such contribution from the continuum ($P$ subspace) should be small, the approximations done in dealing with the escape term $W^\uparrow$ in Eq. (\ref{eq:homega}) (see, e.g., Ref. \cite{Colo1994}) could make an influence and in the end the sum rules given by PVC with the escape term could be slightly different from those of RPA.
The numerical results will be shown in Sec. \ref{sec:res}.

In any case, when the diagonal approximation is removed, the sum rules $m_0$ and $m_1$ will not be influenced as this approximation only affect the interaction $Q_2HQ_2$.
Similar to SRPA, this part will affect the sum rules from $m_3$, which is \cite{Adachi1988}
\begin{align}
  m_3^{\rm SRPA} &= \frac{1}{2} O^\dagger H_{11}^3\mathscr{M} O
  + \frac{1}{2} O^\dagger \left( H_{12}H_{21}H_{11} \right. \notag \\
  &~~~ \left. + H_{11}H_{12}H_{21} + H_{12}H_{22}H_{21} \right) \mathscr{M} O,
\label{eq:}
\end{align}
with expressions $H_{11} = Q_1HQ_1$ and so on.

\section{Numerical details}\label{sec:nd}

The nucleus $^{16}$O is studied as an example since it provides a simple case for various theoretical investigations and tests.
As it is a doubly magic nucleus, the effects of pairing and deformation can be ignored.
The Skyrme functional SAMi \cite{Roca-Maza2012} will be used in all calculations except in the last section where a systematic study on the dependence on the parameterization of the Skyrme functional is given.
Three isoscalar (IS) non charge-exchange excitation modes will be examined: the giant monopole resonance (GMR, $J^\pi = 0^+$), giant dipole resonance (GDR, $J^\pi = 1^-$), and giant quadrupole resonance (GQR, $J^\pi = 2^+$).
The corresponding adopted excitation operators are \cite{Colo2013}
\begin{subequations}\label{eq:}\begin{align}
  O({\rm ISGMR}) &= \sum_{i=1}^A r_i^2 Y_{00}, \\
  O({\rm ISGDR}) &= \sum_{i=1}^A \left( r_i^3 - \frac{5\langle r^2 \rangle}{3} r_i \right) Y_{1M}, \\
  O({\rm ISGQR}) &= \sum_{i=1}^A r_i^2 Y_{2M},
\end{align}\end{subequations}
with $r_i$ the radial coordinate of the $i$'th nucleon and $Y_{LM}$ the spherical harmonic function.
The special form of the ISGDR is aimed at removing the contribution from the spurious state \cite{Colo2013}.
The spurious state in the RPA solution has also been excluded in the selection of doorway states $|N\rangle$ in Eq.~(\ref{eq:doorway}).
The corresponding EWSR is evaluated by the double commutator (DC) in Eq.~(\ref{eq:dbc}) with HF ground state $|0\rangle = |\Phi_0^{\rm HF}\rangle$ \cite{Colo2013}:
\begin{subequations}\label{eq:m1dc}\begin{align}
  m_1^{\rm (DC)}({\rm ISGMR}) &= \frac{\hbar^2}{2m} \frac{A}{\pi} \langle r^2 \rangle, \\
  m_1^{\rm (DC)}({\rm ISGDR}) &= \frac{\hbar^2}{2m} \frac{A}{4\pi} \left( 33\langle r^4 \rangle - 25\langle r^2 \rangle^2 \right), \\
  m_1^{\rm (DC)}({\rm ISGQR}) &= \frac{\hbar^2}{2m} \frac{25A}{2\pi} \langle r^2 \rangle,
\end{align}\end{subequations}
with $m$ the nucleon mass.
To take into account the 1-body center-of-mass correction, in the end the DC sum rules are to be multiplied by a factor of $(A-1)/A$.

\begin{figure*}[htbp!]
  \hspace{-0.5cm}
  \includegraphics[width=6.3cm]{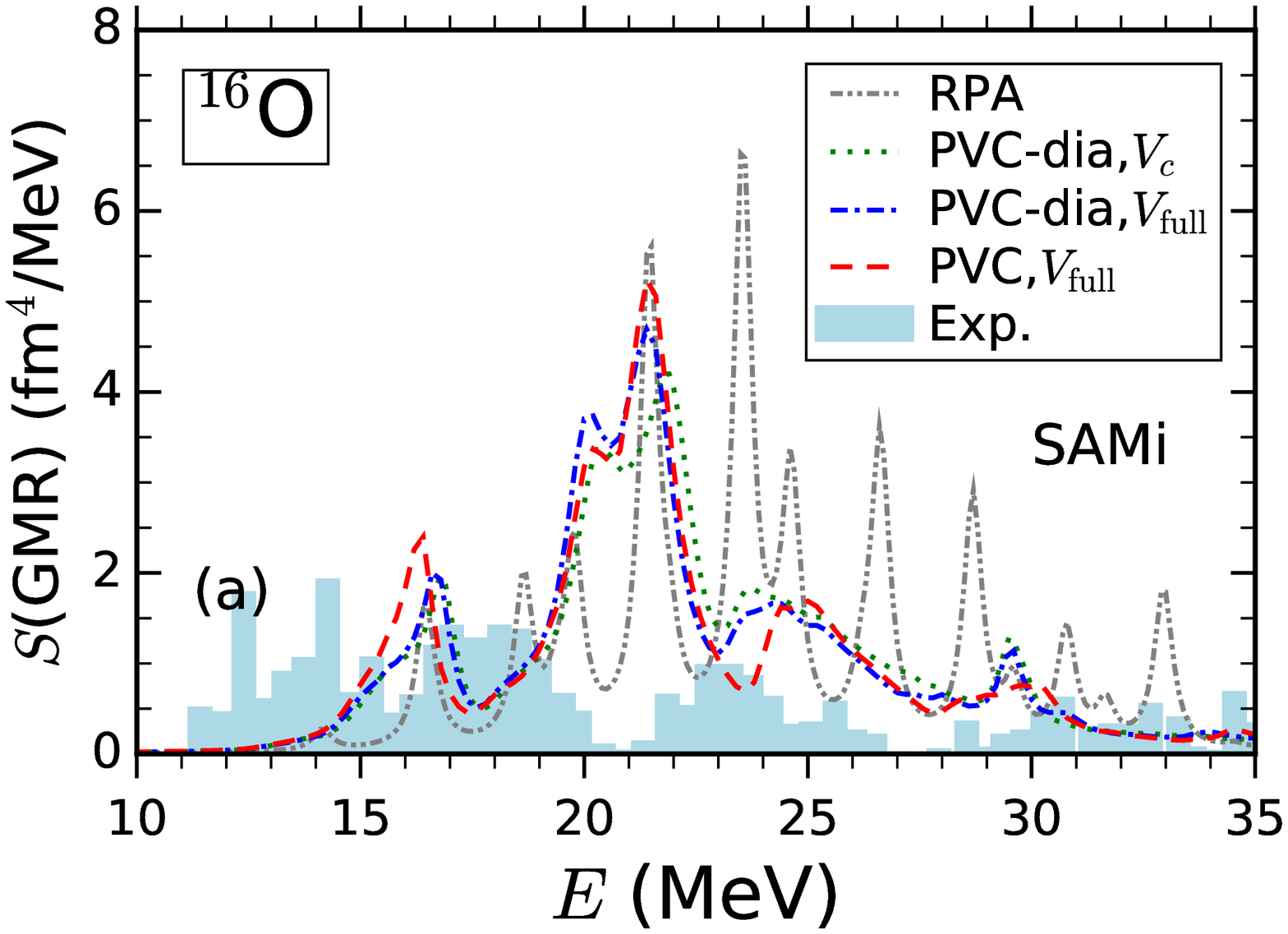}
  \hspace{-0.5cm}
  \includegraphics[width=6.3cm]{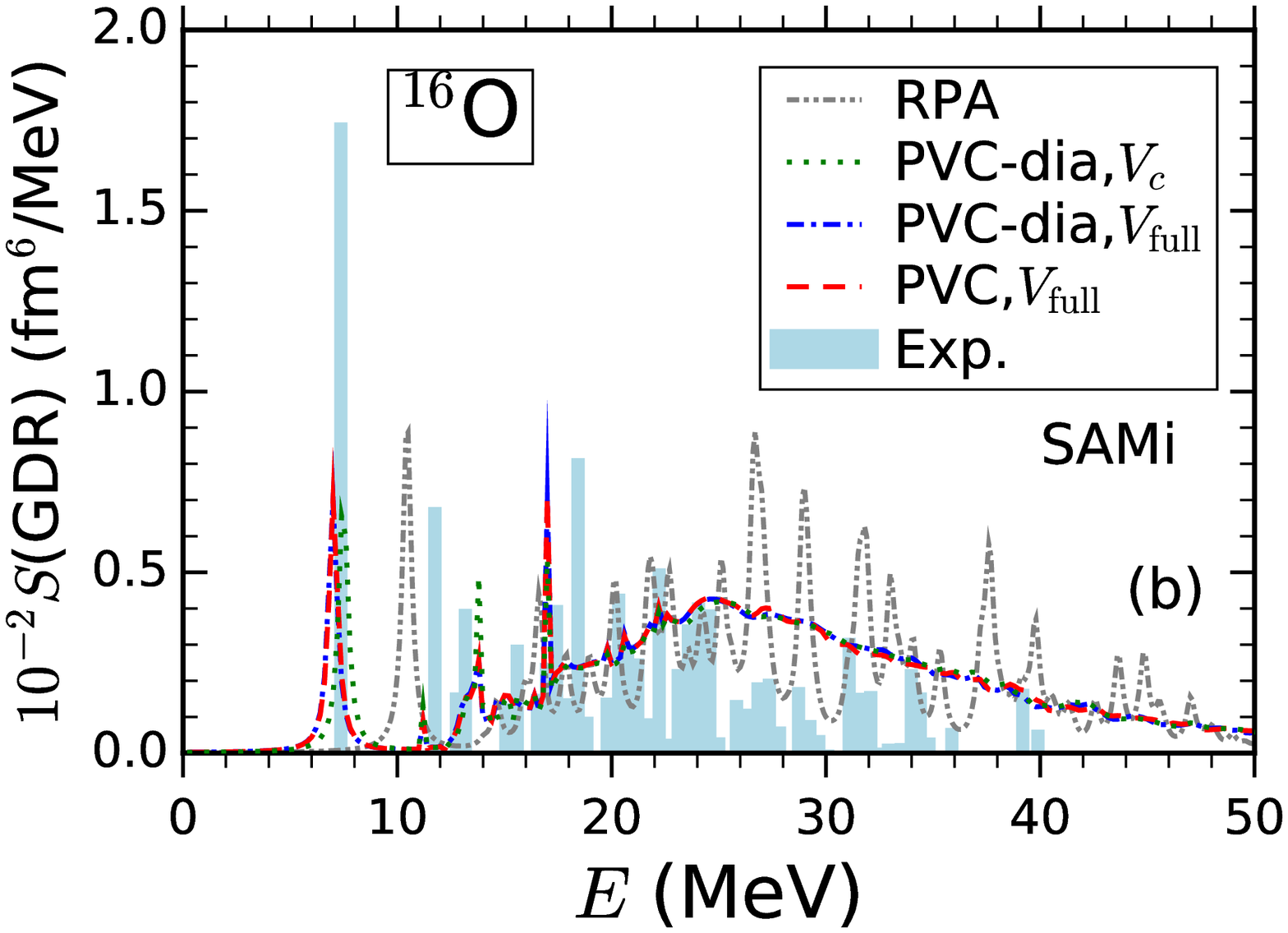}
  \hspace{-0.5cm}
  \includegraphics[width=6.3cm]{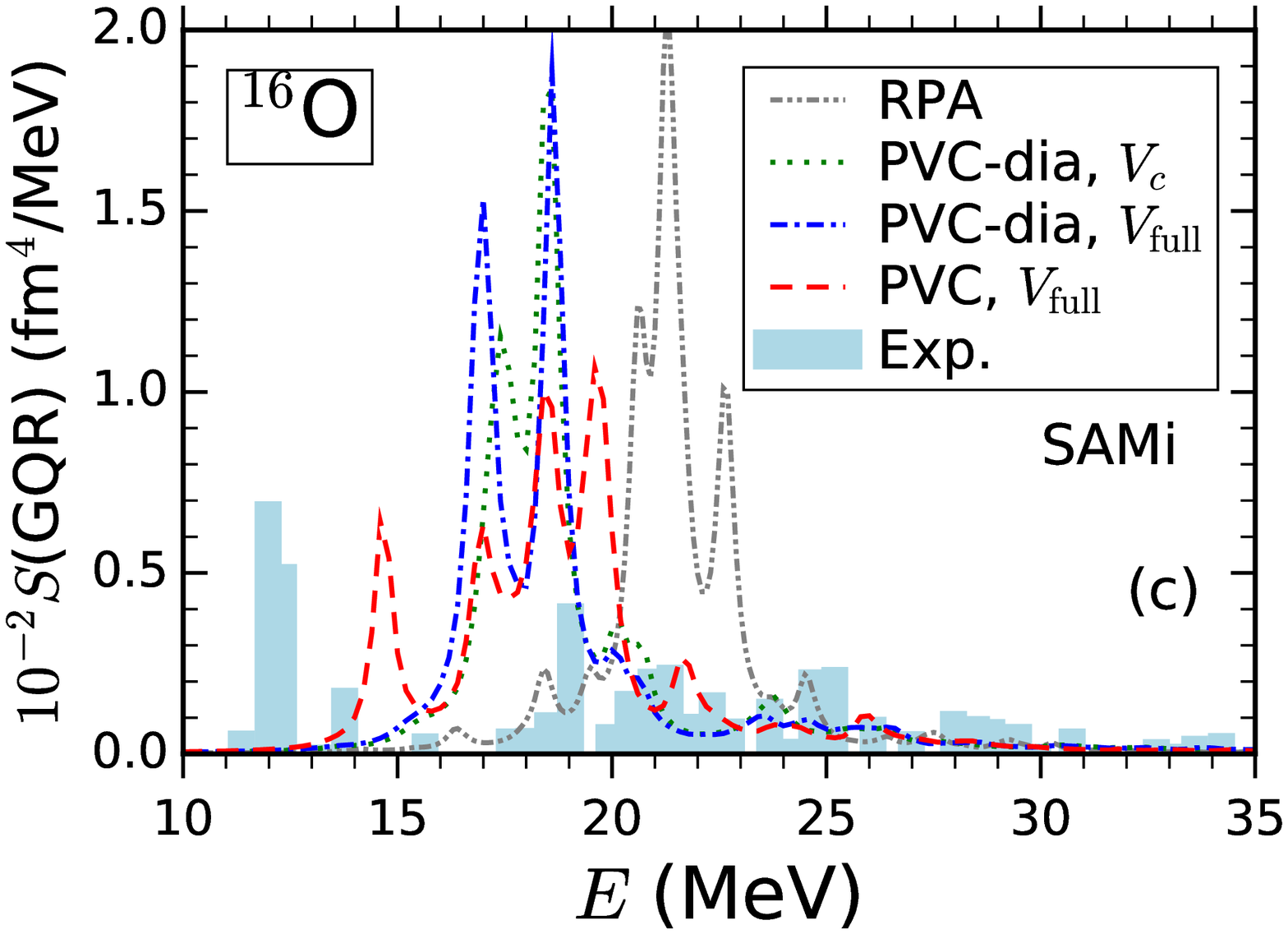}
  \hspace{-0.2cm}
  \caption{(Color online) Strength function of ISGMR (a), ISGDR (b), and ISGQR (c) in $^{16}$O calculated by RPA with full interaction, PVC with diagonal approximation and central interaction (PVC-dia, $V_c$), with diagonal approximation and full interaction (PVC-dia, $V_{\rm full}$), without diagonal approximation and with full interaction (PVC, $V_{\rm full}$).
  In all cases the SAMi functional is used.
  See text for the detail of the experimental data \cite{Harakeh1981,Lui2001}.}
  \label{fig:str}
\end{figure*}

The HF equation is solved in a spherical box with size $R = 20$ fm and a radial step $dr = 0.1$ fm.
In the RPA calculation, the single-particle energy cut-off is $e_{\rm cut} = 80$ MeV, so that it ensures the convergence of our results as it can be seen from the EWSR for ISGMR, ISGDR, and ISGQR  in $^{16}$O that are all $100\%$ fulfilled, see the column ``RPA'' in Table~\ref{tab:sr}.
For the PVC calculation, the phonons selected in the doorway states, i.e., the summation index $n$ in Eq.~(\ref{eq:wph}), include multipolarity $J^\pi = 0^+, 1^-, 2^+, 3^-, 4^+, 5^-$.
Contributions from unnatural parity states such as $0^-$ should be negligible and therefore are not included.
Convergence of the results by considering natural parity phonons up to $5^-$ is well achieved.
The phonon energy cut-off is $\omega_{n,{\rm cut}} = 30$ MeV.
A further criteria for the selection of phonons is its strength, only those phonons with $B(EJ)/m_0 \geq F_{\rm cut}$ will be selected in the doorway states and the fraction cut-off is $F_{\rm cut} = 2\%$.
These cut-offs have been checked in previous investigations \cite{Roca-Maza2017}.
The smearing parameter $\epsilon$ in Eq.~(\ref{eq:homega}) is chosen as $0.25$ MeV.

\section{Results and discussion}\label{sec:res}

\subsection{Spectrum and sum rules}\label{sec:str}

In Fig.~\ref{fig:str} we show the strength function of ISGMR, ISGDR, and ISGQR in $^{16}$O calculated by RPA (bars) and PVC (lines), in comparison with experimental data \cite{Harakeh1981,Lui2001}. The original data is given in terms of the fraction of EWSR $F(E)$ in Ref.~\cite{Lui2001}, with a total of $(48\pm 10)\%$, $(32\pm 7)\%$, and $(53\pm10)\%$ of the EWSR in the region $E_x$ from 11 to 40 MeV. This data is transformed to the strength distribution by:
\begin{equation}\label{eq:fe}
  S(E) = \frac{F(E)}{E} m_1,
\end{equation}
with the values of $m_1$ adopted as the double commutator ones in Table~\ref{tab:sr}.
For the dipole resonance, the level at $7.12$ MeV which exhausts $4.2\%$ of the EWSR is taken from Ref. \cite{Harakeh1981}.

In previous studies of PVC such as Refs.~\cite{Colo1994,Roca-Maza2017}, the interaction vertex $Q_1HQ_2$ in Eq.~(\ref{eq:q1hq2}) includes only the central term of the Skyrme interaction.
The effect of other terms on the single-particle properties have been investigated in Ref.~\cite{Colo2010,CaoLG2014}.
Here we would like to investigate the effect of those terms on the strength function, therefore in Fig.~\ref{fig:str} both the results of PVC with central interaction ($V_c$) and with full interaction ($V_{\rm full}$) are given, within the diagonal approximation (PVC-dia).
For PVC without diagonal approximation (PVC), only the results with full interaction are given. In all cases, the HF+RPA calculations are performed with full Skyrme interaction.

It can be seen from Fig.~\ref{fig:str} that by including the escape and spreading effects within the PVC, the width of the strength distribution appears naturally, unlike in the case of RPA. This makes the comparison with experimental data more realistic. On the other hand, the centroid of the distribution ($m_1/m_0$) is shifted to a lower energy, from few hundreds of keV for the ISGMR and ISGDR to a maximum of about 1.5 MeV for the case of the ISGQR (cf. Table \ref{tab:sr}). It is important to note here that functionals are usually calibrated in order to give a reasonable description of the experimental centroid energy at the RPA level and, therefore, such shift may lead to worse agreement with the data.

By comparing the results with central term only and results with full interaction, it can be seen that by including Coulomb term and spin-orbit term, the strength is generally slightly shifted to a lower energy. In the case of ISGMR and ISGDR, the shape of the strength distribution does not change too much, while in ISGQR such change is more significant.

From PVC-dia to PVC, the strength function is also much influenced in the ISGQR case. For PVC-dia, there are two major peaks near 17 and 18.5 MeV; while for PVC, there are four major peaks near 14.5, 17, 18.5, and 19.5 MeV, with lower strength and wider distribution. The lowest peak near 14.5 MeV is of particular interest as there is no sign of this peak in PVC-dia. It will be used as an example in Section \ref{sec:dia} to analyze the difference between calculation with and without the diagonal approximation. Regarding the ISGMR, the removal of the diagonal approximation also shows some effect, for example: the lowest peak near 16.5 MeV is slightly shifted to a lower energy and the strength increases; The distributions of the peaks near 20, 21.5, 24, and 29.5 MeV are also affected, but, overall, the effect is weaker than the case of ISGQR. Among the three cases, the ISGDR is the one where the diagonal approximation shows less influence.

\begin{figure*}[htbp]
  \hspace{-0.5cm}
  \includegraphics[width=6.3cm]{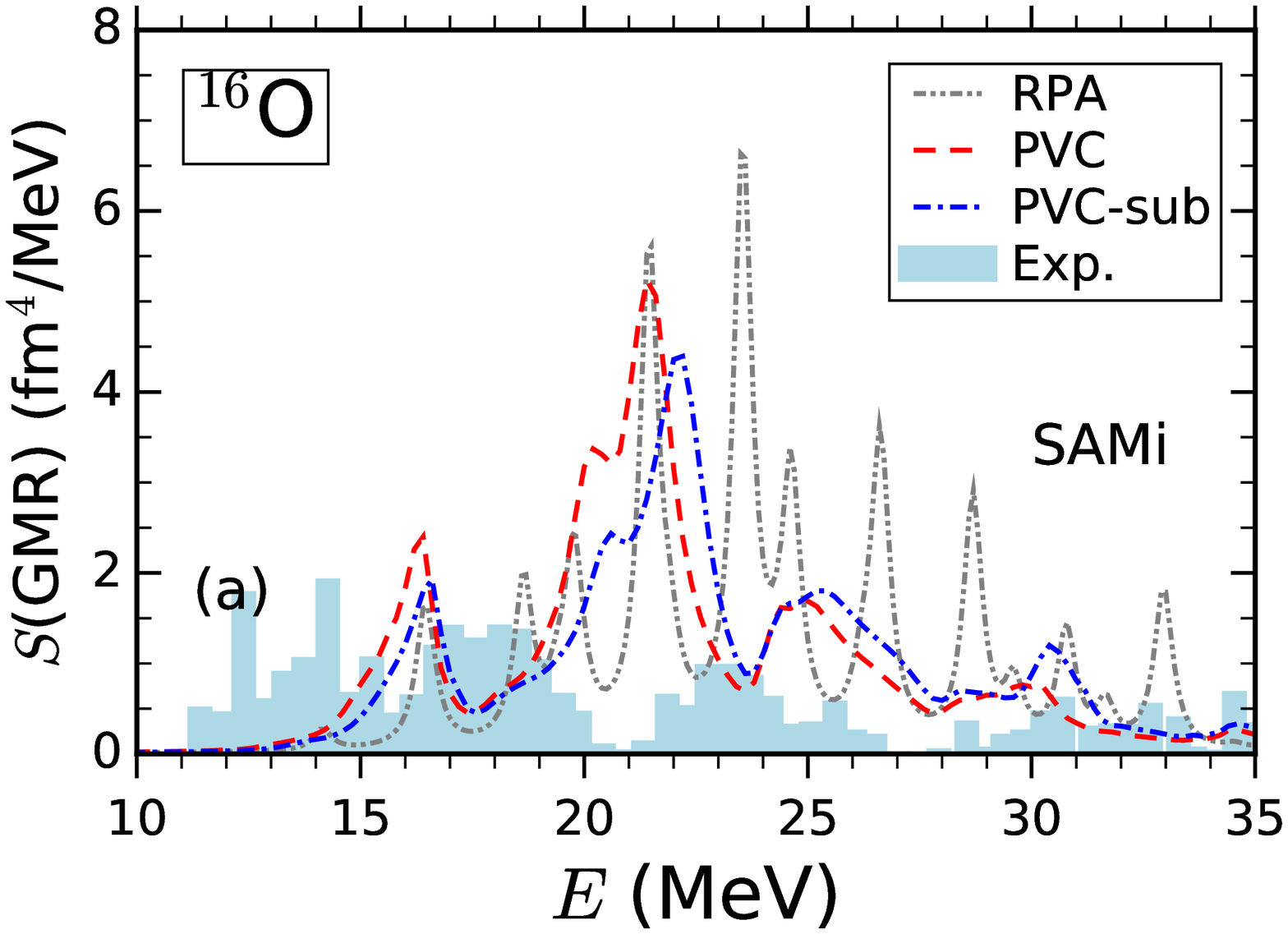}
  \hspace{-0.5cm}
  \includegraphics[width=6.3cm]{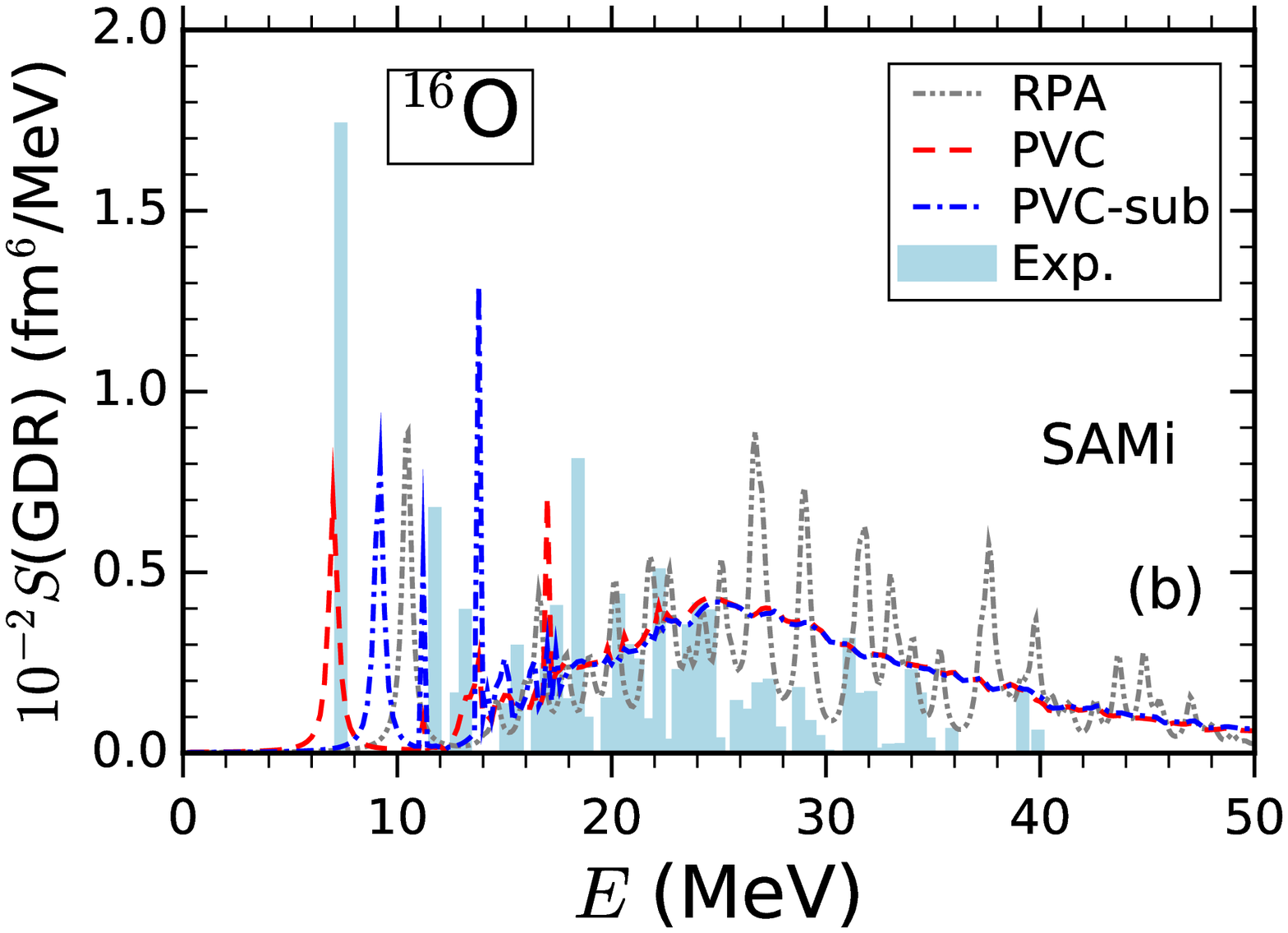}
  \hspace{-0.5cm}
  \includegraphics[width=6.3cm]{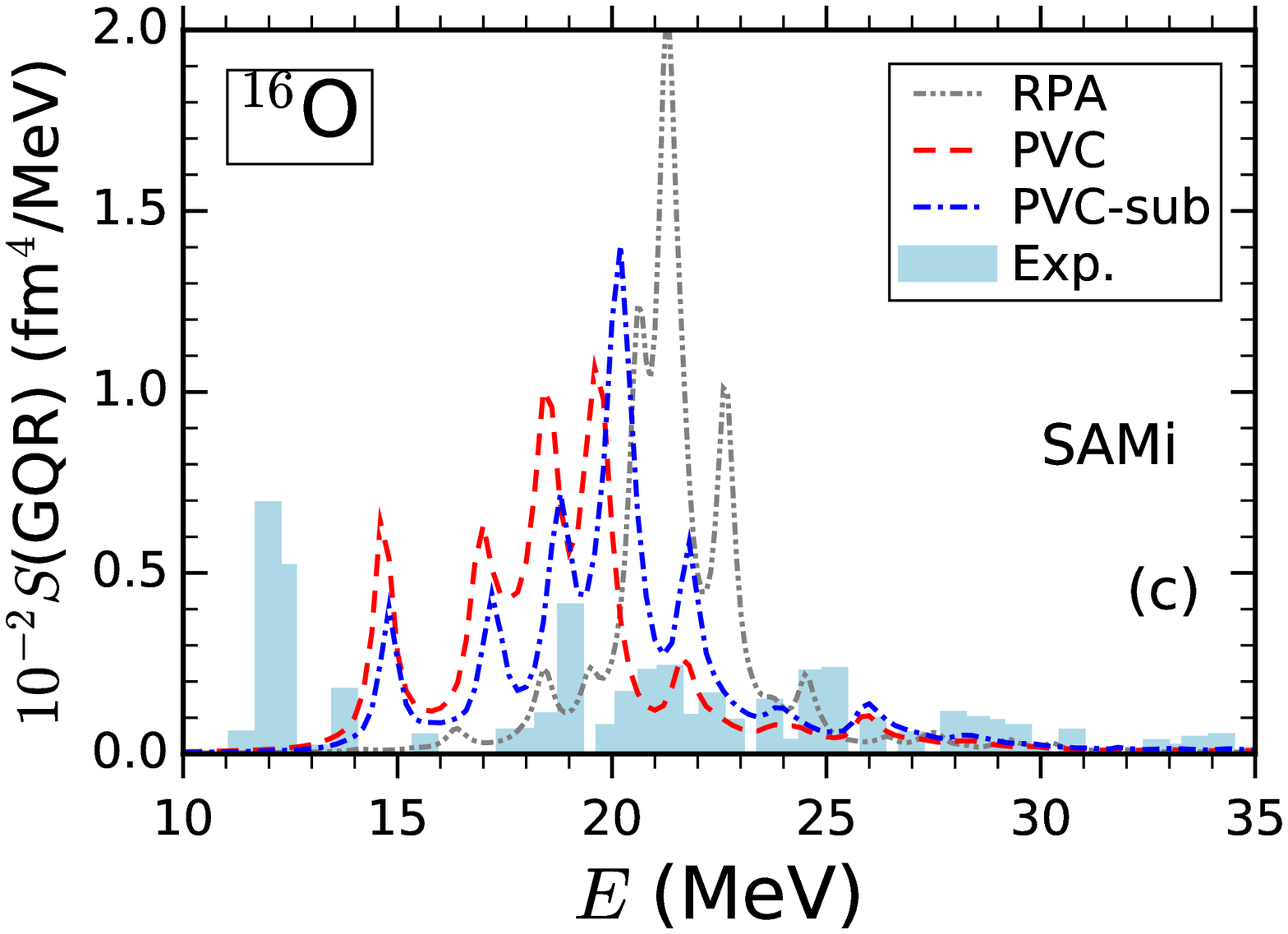}
  \hspace{-0.2cm}
  \caption{(Color online) Same as Fig.~\ref{fig:str}, but for comparison between PVC calculation with and without subtraction.}
  \label{fig:str2}
\end{figure*}

As mentioned in the Introduction, the diagonal approximation has been investigated in the SRPA framework for the giant resonances of $^{16}$O in Ref.~\cite{Gambacurta2010}.
From RPA to SRPA, the strength distributions are shifted towards a lower energy, similar to the effect of PVC in Fig.~\ref{fig:str}.
However, quantitatively, the effect of SRPA is larger.
For ISGMR, IVGDR, and ISGQR, the main peaks are shifted towards a lower energy by about 4, 6, and 8 MeV, respectively ~\cite{Gambacurta2010}; while for PVC the shifts are $\approx 2-3$ MeV.
For the ISGMR, the diagonal approximation in SRPA shifts the distribution to a lower energy by about 2 MeV, while in PVC it changes mildly (see Fig.~\ref{fig:str} (a) of this work and Fig. 8 (a) of Ref.~\cite{Gambacurta2010}).
For the dipole case, the effect of diagonal approximation is small in both SRPA and PVC (see Fig.~\ref{fig:str} (b) of this work and Fig. 15 (b) of Ref.~\cite{Gambacurta2010}).
For the ISGQR, the diagonal approximation in SRPA shifts the distribution towards a lower energy by around 2 MeV, similar to ISGMR, while in PVC it is more complicated as the shape has changed much (see Fig.~\ref{fig:str} (c) of this work and Fig. 9 (a) of Ref.~\cite{Gambacurta2010}).
In all the cases, the diagonal approximation in SRPA does not change much the shape of the strength distribution, while in PVC this is not the case for the ISGQR.

The diagonal approximation has also been studied in the RTBA framework \cite{Litvinova2010,Litvinova2013}, where it was removed by including additional phonon coupling between two quasiparticles inside the 2 quasiparticles $\otimes$ phonon configuration.
The low-lying dipole excitations in $^{116,120}$Sn and $^{68,70,72}$Ni were investigated.
By removing the diagonal approximation, a larger fraction of the pygmy mode is pushed above the neutron threshold.
For a detailed comparison between RTBA and current framework, see Appendix \ref{app:tba}.

In Fig.~\ref{fig:str2} the effects of the subtraction [Eq.~(\ref{eq:sub})] in the PVC calculation are shown for the ISGMR, ISGDR, and ISGQR strength distributions of $^{16}$O. As a reference, the results of the RPA and experimental data shown in Fig.~\ref{fig:str} are also displayed in Fig.~\ref{fig:str2}. It can be seen that by adopting the subtraction procedure, the strength distributions are generally shifted towards a higher energy by about 1 MeV, except in the ISGDR case where the main peak at 17 MeV vanishes and a new peak at 14 MeV appears.
The effects of subtraction presented here are consistent with the findings of previous investigation using PVC-dia (see Fig. 4 of Ref.~\cite{Roca-Maza2017}).

In Ref.~\cite{Lyutorovich2015}, the quasiparticle-phonon coupling model with time-blocking approximation was used to study the ISGMR, ISGQR, and isovector GDR of $^{16}$O, $^{40}$Ca, and $^{208}$Pb.
A systematic downward shift of the centroid energy of the giant resonances was found from RPA to TBA with subtraction.
This effect is similar to the one of PVC presented here, though quantitatively it is smaller (see Fig.~\ref{fig:str2} of this work and Fig. 4 of Ref.~\cite{Lyutorovich2015}). Especially in the case of ISGQR, the PVC calculation (with or without subtraction) gives very different strength distribution from the one given by RPA, while they are similar for TBA and RPA \cite{Lyutorovich2015}.
This might be related with the diagonal approximation as removing it shows quite some effect here.

The subtraction in SRPA has been investigated for ISGMR and ISGQR and it also pushes the strength distribution to a higher energy \cite{Gambacurta2015}.
However, comparing with the results of PVC shown in Fig.~\ref{fig:str2}, the effect in SRPA is again larger.
With subtraction, the strength are shifted towards a higher energy by about 2 MeV in SRPA while in PVC it is generally less than 1 MeV, see Figs. 1 and 4 of Ref.~\cite{Gambacurta2015}.
Comparing the results of SRPA including subtraction with RPA given in Ref.~\cite{Gambacurta2015}, the main peaks of ISGMR and ISGQR given by SRPA with subtraction are about 1.5 and 1 MeV lower than those by RPA.
These are similar to the differences between PVC with subtraction and RPA shown in Fig.~\ref{fig:str2}.

In comparison with the experimental data, the three peaks of ISGMR around 18, 23, and 26 MeV may correspond to the three peaks given by the PVC, though the energies are slightly lower than the data.
This may be understood as the SAMi functional has been developed in such a way that the experimental ISGMR is reproduced at the RPA level, which can be seen from the black vertical lines in Fig.~\ref{fig:str2}.
When the PVC is included, though the description of resonance width has been improved, the centroid is pushed to a slightly lower energy and the subtraction remedies to this problem only to some extent.
The peaks around 12 and 14 MeV may be due to $\alpha$-clustering effects \cite{Yamada2012}.

In the case of ISGDR the description of PVC with SAMi functional is rather good, especially the low-lying 7.1 MeV level has been nicely reproduced.
The peaks around 12 and 18 MeV, and the resonance shape above 20 MeV are also well described.
The subtraction worsens the description of the data below 20 MeV.

In the case of the ISGQR, the peak at 15 MeV by PVC might be attributed to the peak at 12 MeV or 14 MeV of the data.
The experimental data for the high energy part of ISGQR is concentrated from 18 to 26 MeV, while the theoretical distribution is from 16 to 22 MeV, slightly lower than the data. Again, the strength distribution given by subtraction method is shifted towards higher energy, but the peak position is still lower than the experimental data.
It has been shown that the dominant decay channel of ISGQR of $^{16}$O is $\alpha$ emission to the ground and first excited states of $^{12}$C \cite{Knopfle1978}.
Such effect is not included in current PVC framework and could be part of the reasons for the disagreement with the data.
Besides that, taking into account the coupling with more phonons \cite{Litvinova2018a} and ground-state correlation might also help to improve the descriptions of the ISGMR and ISGQR of $^{16}$O.

The sum rules for the above discussed calculations are shown in Table \ref{tab:sr}, including the results for the RPA, PVC-dia without Coulomb and spin-orbit interactions in the PVC vertex (PVC-d, $V_c$), PVC-dia with full interaction (PVC-d), PVC, and PVC with subtraction (PVC-s).
The strength function given by PVC are integrated up to $E = 120$ MeV.
The EWSR $m_1$ by the double commutator in Eq.~(\ref{eq:m1dc}) are also shown.

\begin{figure}[htbp]
  \includegraphics[width=8cm]{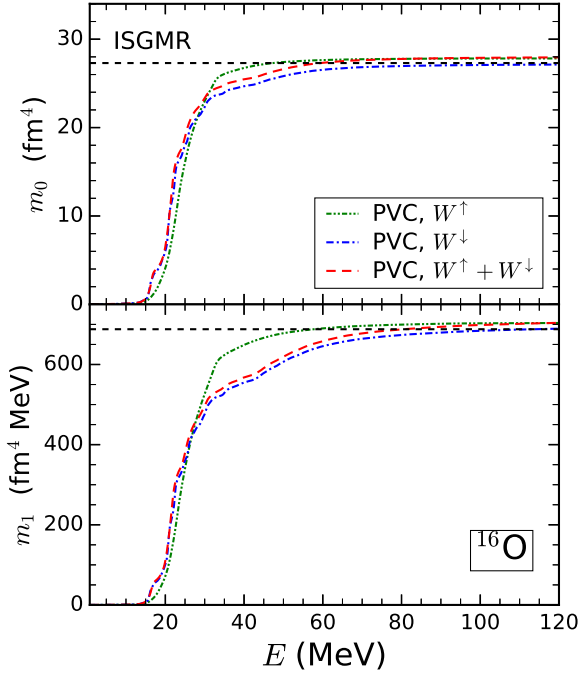}
  \caption{(Color online) Sum rules $m_0$ and $m_1$ of ISGMR in $^{16}$O as a function of the upper limit of the integrated energy range, calculated by PVC with escape term ($W^\uparrow$), spreading term ($W^\downarrow$), and both terms.
  Horizontal black dashed lines are RPA results (for $m_0$) and DC results (for $m_1$), see Table \ref{tab:sr} for the values.}
  \label{fig:sr}
\end{figure}

\begin{table}[h]
  \caption{Sum rules for the ISGMR, ISGDR and ISGQR responses in $^{16}$O calculated by: RPA, PVC-dia without Coulomb and spin-orbit interactions (PVC-d, $V_c$), PVC-dia with full interaction (PVC-d), PVC, PVC with subtraction (PVC-s).
  The EWSR $m_1$ by double commutator (DC) is also given.
  In all cases the SAMi functional is used.
  The units of $m_{-1}, m_0$, and $m_1$ are fm$^4$/MeV, fm$^4$, and fm$^4$ MeV, respectively, for the ISGMR and ISGQR; they are 
  fm$^6$/MeV, fm$^6$, and fm$^6$ MeV for the ISGDR.
  }
  \label{tab:sr}
  \centering
  \begin{ruledtabular}
  \begin{tabular}{l|lcccccc}
  & SR & RPA & DC & PVC-d,$V_c$ & PVC-d & PVC & PVC-s \\
  \hline
  ISGMR    & $m_{-1}$         & 1.14  &     & 1.25 & 1.25  & 1.25  & 1.17  \\
           & $m_0$            & 27.3  &     & 27.9 & 27.8  & 27.8  & 27.8  \\
           & $m_1$            & 689   & 688 & 701  & 701   & 700   & 740   \\
        & $m_{1}/m_0$         & 25.3  &     & 25.2 & 25.2  & 25.2  & 26.6  \\
&$\sqrt{\frac{m_{1}}{m_{-1}}}$& 24.6  &     & 23.7 & 23.7  & 23.7  & 25.1  \\
  \hline
  ISGDR    & $m_{-1}$         & 38.4  &     & 42.5 & 43.0  & 43.0  & 42.8  \\
           & $m_0$            & 968   &     & 981  & 982   & 981   & 1008  \\
           & $m_1$            & 29567 &29493& 29583& 29619 & 29591 & 30607 \\
        & $m_{1}/m_0$         & 30.5  &     & 30.2 & 30.2  & 30.2  & 30.4  \\
&$\sqrt{\frac{m_{1}}{m_{-1}}}$& 27.7  &     & 26.4 & 26.2  & 26.2  & 26.7  \\
  \hline                                                           
  ISGQR    & $m_{-1}$         & 18.4  &    & 22.6 & 23.2  & 23.3  & 20.9  \\
           & $m_0$            & 397   &    & 419  & 420   & 420   & 413   \\
           & $m_1$            & 8613  &8604& 8503 & 8488  & 8489  & 8949  \\
        & $m_{1}/m_0$         & 21.7  &    & 20.3 & 20.2  & 20.2  & 21.7  \\
&$\sqrt{\frac{m_{1}}{m_{-1}}}$& 21.6  &    & 19.4 & 19.1  & 19.1  & 20.7  \\
  \end{tabular}
  \end{ruledtabular}
\end{table}
First, the EWSR ($m_1$) values given by the RPA calculation in all three cases, the ISGMR, ISGDR, and ISGQR, are fully exhausted comparing with the ones obtained by the double commutator.
For the PVC results, there are small discrepancies.
As discussed in Sec. \ref{sec:sr}, when only spreading term is taken into account, the $m_0$ and $m_1$ given by PVC should be the same as those of RPA, similar to the case of SRPA \cite{Adachi1988}.
To verify this, we show in Fig.~\ref{fig:sr} the $m_0$ and $m_1$ as a function of the upper limit of the integrated energy range.
The case of ISGMR is taken as an example, while the others give similar results.
In this figure the horizontal dashed line is the RPA results for $m_0$ and DC for $m_1$ (see Table \ref{tab:sr}).
One can see that when only spreading term ($W^\downarrow$) is included in the PVC calculation, the $m_0$ and $m_1$ are the same as those of RPA.

When escape term ($W^\uparrow$) is included, the sum rules $m_0$ and $m_1$ are slightly different from those of RPA.
This could be due to the approximation in the escape term that the interaction has not been taken into account \cite{Colo1994}.
To verify this, we compare the sum rules of PVC with escape term only starting from RPA phonons, to the results that starting from unperturbed phonons (without particle-hole interactions).
The latter one is labeled UNP and the results are listed in Table \ref{tab:sr2}.
To achieve a higher precision, the smearing parameter $\epsilon$ in Eq.~(\ref{eq:homega}) is chosen as $0.1$ MeV in this Table instead of 0.25 MeV in other calculations.
It can be seen that when interactions are not taken into account at the beginning, the $m_0$ and $m_1$ given by PVC with escape term are almost the same as original calculation in $Q_1$ subspace, with a relative difference 0.2\%.
When interaction is included in $Q_1$ subspace (RPA calculation) but not in $P$ subspace (PVC with escape term), the sum rules will slightly be influenced.

\begin{table}[h]
  \caption{Sum rules for the ISGMR in $^{16}$O calculated by RPA (or unperturbed calculation, UNP) and PVC with escape term only $(W^\uparrow)$.
  The relative differences between PVC and RPA (or UNP) are also given $(\delta)$.
  In all cases the SAMi functional is used.
  }
  \label{tab:sr2}
  \centering
  \begin{ruledtabular}
  \begin{tabular}{l|ccc|ccc}
  SR & RPA & $W^\uparrow$ & $\delta$ & UNP & $W^\uparrow$ & $\delta$ \\
  \hline
  $m_0$ (fm$^4$)    & 27.27 & 27.88 & 2.2\% & 28.13 & 28.06 & 0.2\% \\
  $m_1$ (fm$^4$MeV) & 689.1 & 702.8 & 2.0\% & 816.8 & 815.5 & 0.2\% \\
  \end{tabular}
  \end{ruledtabular}
\end{table}

As it has been discussed in Fig.~\ref{fig:str}, the strength distributions given by PVC are generally shifted to a lower energy comparing with those by RPA . Therefore, the inverse EWSR ($m_{-1}$) are larger compared with RPA even when the subtraction method is implemented, and the centroid energy ($m_1/m_0$ or $\sqrt{m_1/m_{-1}}$) are smaller. The influence of non-central terms of the interaction (comparing PVC-d,$V_c$ and PVC-d) on the sum rules are negligible in the case of ISGMR, while for ISGDR and ISGQR a small effect shows up.
In all cases, the diagonal approximation (comparing PVC-d and PVC) has little influence on the sum rules.
On the other hand, the subtraction has much influence on the sum rules (comparing PVC and PVC-s).
The EWSR are significantly larger and agree less with the double commutator sum rule when subtraction is performed, in agreement with the findings in the PVC-dia calculation \cite{Roca-Maza2017} and SRPA \cite{Gambacurta2015}. This is a feature of the subtraction method that needs to be better investigated. We recall here that the subtraction method was devised for exactly keeping the $m_{-1}$ value obtained within the RPA in beyond RPA calculations while no procedure of renormalization was imposed on $m_1$.

\subsection{Different components of interaction}

Next, we show how different components of the interaction contribute to the strength function in Fig.~\ref{fig:str}, using the ISGQR as an example.

\begin{figure}[htbp]
  \includegraphics[width=8cm]{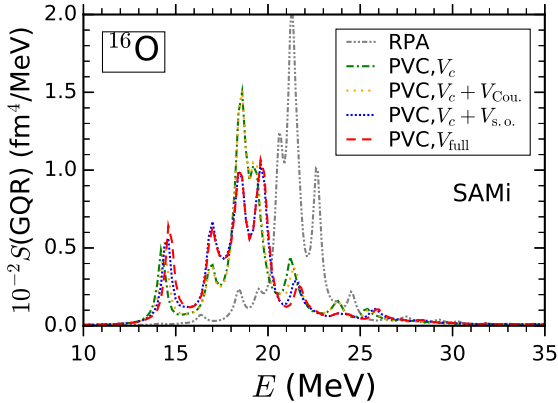}
  \caption{(Color online) Strength function of ISGQR in $^{16}$O calculated by RPA with full interaction, PVC with central interaction (PVC, $V_c$), PVC with central plus Coulomb interactions (PVC, $V_c + V_{\rm Cou.}$), PVC with central plus spin-orbit interactions (PVC, $V_c + V_{\rm s.o.}$), and PVC with full interaction (PVC, $V_{\rm full}$).
  In all cases the SAMi functional is used.}
  \label{fig:strvso}
\end{figure}

In Fig.~\ref{fig:strvso} the strength distributions calculated by PVC and with different terms of interaction are shown, including: with central terms only ($V_c$), with central terms and Coulomb term ($V_{c}+V_{\rm Cou.}$), with central terms and spin-orbit term ($V_c+V_{\rm s.o.}$), and with full interaction ($V_{\rm full}$).
It can be seen that the Coulomb interaction has a negligible effect on the strength distribution, except for a small influence near 14 and 21 MeV.
On the other hand, the spin-orbit term has much influence and clearly changes the distribution.
With the central term there is only one minor peak near 14 MeV and one major peak near 18.5 MeV.
When including the spin-orbit term, the strength of the major peak decreases much and two other peaks near 17 and 19.5 MeV become larger.

\subsection{Influence of diagonal approximation: eigen-energies}\label{sec:dia}

In this Subsection, we analyze the difference between PVC with and without diagonal approximation in Fig.~\ref{fig:str}.
The low energy peak at $\omega = 14.6$ MeV in the ISGQR will be used as an example, as it 
is manifestly different in the two calculations.
In the following results, all PVC calculations are performed with the full interaction at $\omega = 14.6$ MeV.
The integration of the strength around this energy ($14.6 \pm 0.4$ MeV) within PVC calculation gives 
$\int S_{\rm PVC} d\omega = 34.9$ fm$^4$, whereas within PVC-dia is $\int S_{\text{PVC-dia}} d\omega = 3.7$ fm$^4$.

\begin{figure}[htbp]
  \includegraphics[width=8cm]{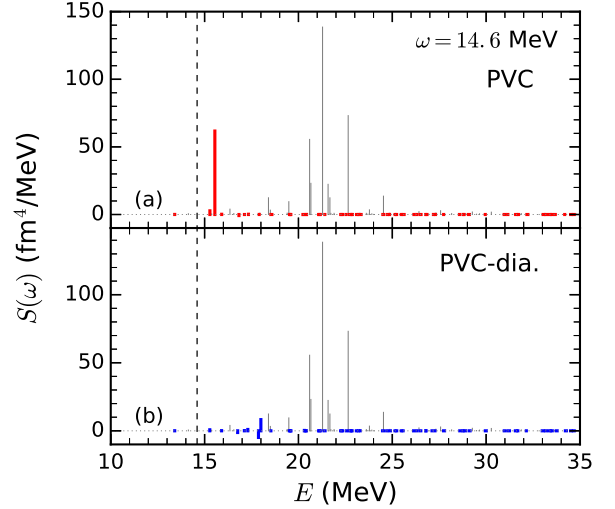}
  \caption{(Color online) Contribution to the ISGQR strength function of $^{16}$O at $\omega = 14.6$ MeV from different PVC eigenstates $\nu$ with different excitation energies, see also Eq.~(\ref{eq:str}).
  Results of (a) PVC (red lines) and (b) PVC-dia (blue lines) are shown.
  The position of $\omega = 14.6$ MeV is given by the vertical dashed line, and the RPA states are given by the gray vertical lines (with unit fm$^4$).}
  \label{fig:strn}
\end{figure}

Figure \ref{fig:strn} shows the strength contributions from different PVC states $\nu$, as given by Eq.~(\ref{eq:str}).
The position of $\omega = 14.6$ MeV has been indicated by the vertical dashed line, and the RPA states are given by the gray vertical lines in the background.
For PVC, the largest contribution to the strength at $\omega = 14.6$ MeV comes from the PVC state at $\Omega_\nu = 15.5$ MeV; while for PVC-dia, the largest contribution comes from the state at $\Omega_\nu = 18.0$ MeV.

\begin{figure}[htbp]
  \includegraphics[width=8cm]{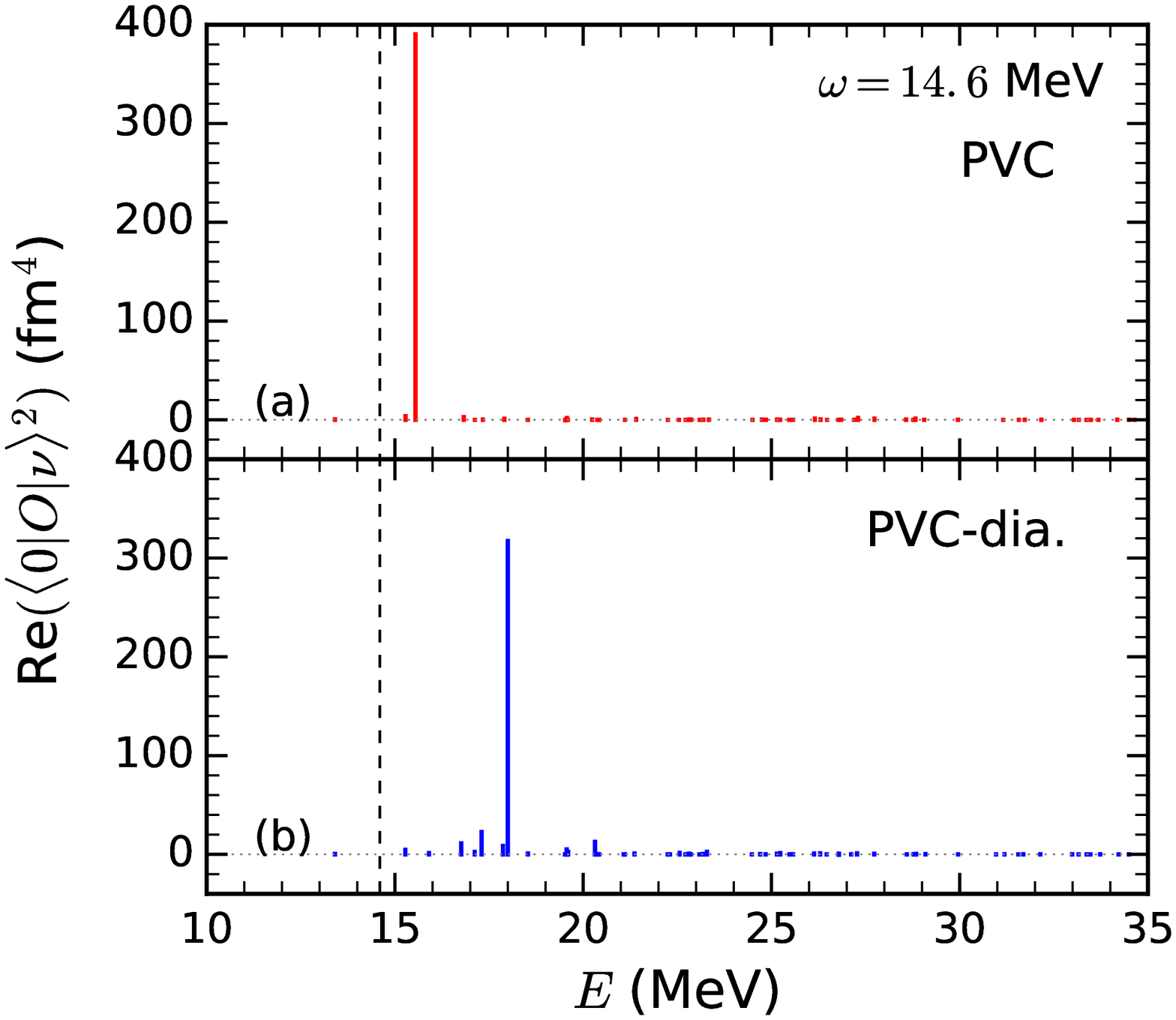}
  \caption{(Color online) Real part of the square of the transition matrix element $\langle 0|O|\nu \rangle$ for PVC states $\nu$ with different excitation energies.
  The calculation is performed for the ISGQR of $^{16}$O at $\omega = 14.6$ MeV (vertical dashed line) by (a) PVC and (b) PVC-dia.}
  \label{fig:o0n}
\end{figure}

Let us express the square of the transition matrix elements explicitly in terms of its real and imaginary parts,
\begin{equation}\label{eq:}
  \langle 0|O|\nu \rangle^2 = a_\nu + ib_\nu,
\end{equation}
with $a_\nu$ and $b_\nu$ both real numbers.
From Eq.~(\ref{eq:str}), the strength function can be written as
\begin{equation}\label{eq:str2}
  S(\omega) = \frac{1}{\pi} \sum_\nu \frac{a_\nu\frac{\Gamma_\nu}{2}-b_\nu(\omega-\Omega_\nu)}{(\omega-\Omega_\nu)^2+\frac{\Gamma_\nu^2}{4}}.
\end{equation}
In Fig.~\ref{fig:o0n} 
the real part of the square of the transition matrix element $a_\nu =$ Re($\langle 0|O|\nu \rangle^2$) is shown.
It can be seen that both the transition matrix element 
of the state at $\Omega_\nu = 15.5$ MeV in PVC, 
and that of the state at $\Omega_\nu = 18.0$ MeV in PVC-dia, are very large.
Although the value of the $\Omega_\nu({\rm PVC}) = 15.5$ MeV one is slightly larger than the one of the $\Omega_\nu(\text{PVC-dia}) = 18.0$ MeV, the difference is not large enough to explain the difference in the final contribution to the strength shown in Fig.~\ref{fig:strn}.
Therefore, according to Eq.~(\ref{eq:str2}), the much stronger strength in the PVC from $\Omega_\nu({\rm PVC}) = 15.5$ MeV state must be due to the position of this state, which is much closer to the energy being evaluated, that is, $\omega = 14.6$ MeV.
In this way, the energy denominator in Eq.~\ref{eq:str2} of this state is much smaller than the $\Omega_\nu(\text{PVC-dia}) = 18.0$ MeV state and as a consequence the strength is larger.

Next, we will study the origin of the large difference in the eigenenergies of these two states.
First, one needs to identify the components of these two states, or more specifically, from which RPA states they come from.
For this purpose, we will identify them by looking at the corresponding wave functions.

\begin{figure}[htbp]
  \includegraphics[width=8cm]{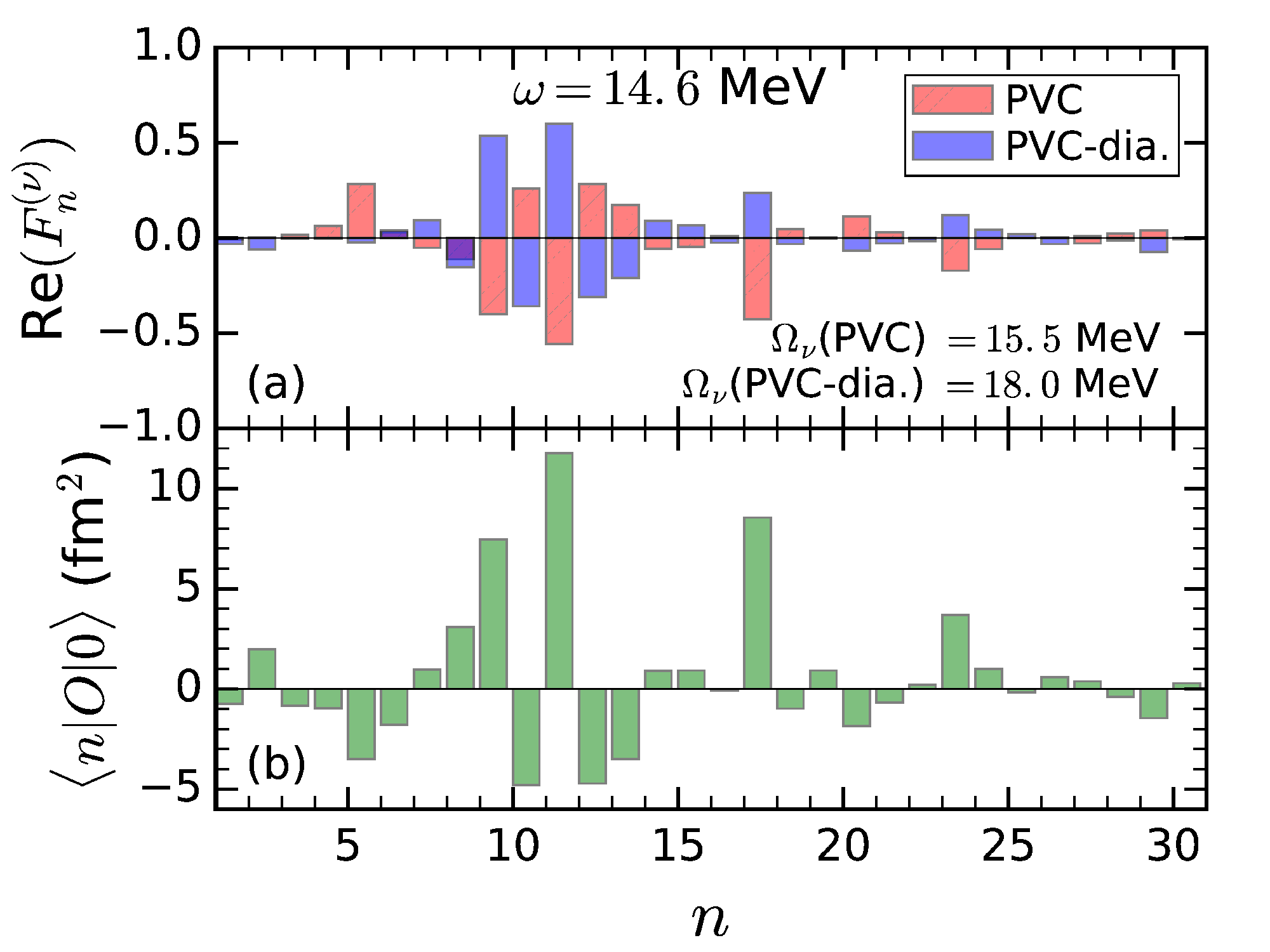}
  \caption{(Color online) (a) Real part of the PVC wave function (\ref{eq:wf}) of states $\Omega_\nu({\rm PVC}) = 15.5$ MeV and $\Omega_\nu(\text{PVC-dia}) = 18.0$ MeV in the basis of RPA states $|n\rangle$ by PVC and PVC-dia.
  (b) Transition matrix elements $\langle n|O|0 \rangle$.}
  \label{fig:wf}
\end{figure}

In the upper panel of Fig.~\ref{fig:wf} we show the real part of the PVC wave function $F_n^{(\nu)}$ (\ref{eq:wf}) of states $\Omega_\nu({\rm PVC}) = 15.5$ MeV and $\Omega_\nu(\text{PVC-dia}) = 18.0$ MeV in the basis of RPA states $|n\rangle$ by PVC and PVC-dia.
The transition matrix elements $\langle n|O|0 \rangle$ in the RPA representation are shown in the lower panel.
The transition matrix elements $\langle 0|O|\nu \rangle$ in the PVC representation in Fig.~\ref{fig:o0n} can be calculated as
\begin{equation}\label{eq:0on}
  \langle 0|O|\nu \rangle = \sum_n F_n^{(\nu)} \langle 0|O|n \rangle.
\end{equation}

It can be seen that these two states have similar RPA components in both PVC calculations.
Although not really dominant, the major component of these two states can be identified as the 11th RPA state with the largest transition matrix element $\langle n=11|O|0 \rangle$.
This RPA state is the one located at $\omega_n = 21.3$ MeV with the largest strength as shown in Fig.~\ref{fig:str} (c) or Fig.~\ref{fig:strn}.

\begin{figure}[htbp]
  \includegraphics[width=8cm]{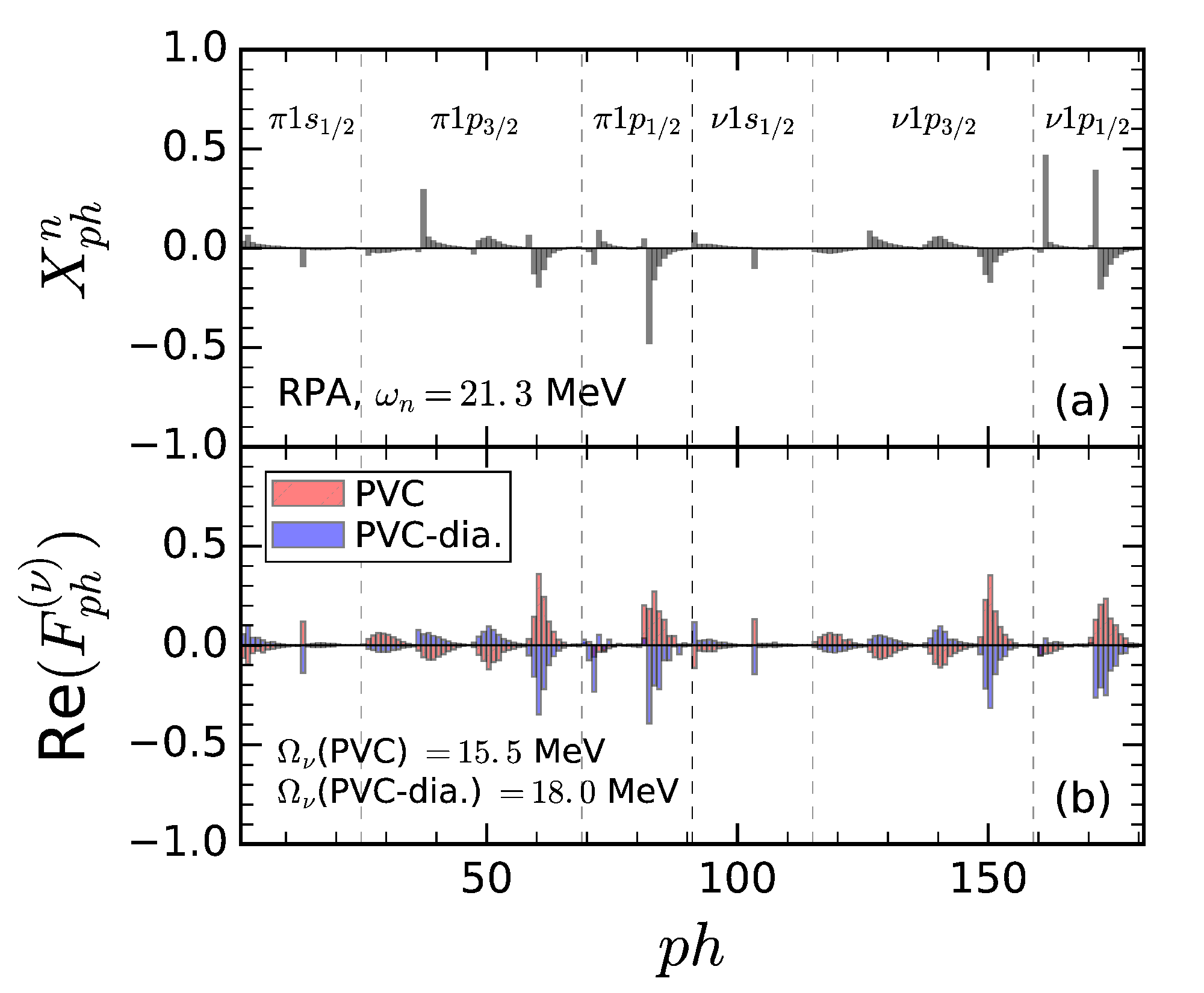}
  \caption{(Color online) (a) RPA wave function ($X$ amplitude) of state with excitation energy $\omega_n = 21.3$ MeV in the 1p-1h representation, see Eq.~(\ref{eq:qn}).
  Different regions divided by vertical lines are for different hole states.
  (b) Real part of the PVC wave functions of states with excitation energies $\Omega_\nu({\rm PVC}) = 15.5$ MeV and $\Omega_\nu(\text{PVC-dia}) = 18.0$ MeV, see Eq.~(\ref{eq:fph}), by PVC and PVC-dia.}
  \label{fig:wfph}
\end{figure}

In Fig.~\ref{fig:wfph}, the wave function $F_n^{(\nu)}$ of states $\omega_\nu({\rm PVC}) = 15.5$ MeV and $\omega_\nu(\text{PVC-dia}) = 18.0$ MeV are transformed to the 1p-1h basis for the $X$ amplitude as
\begin{equation}\label{eq:fph}
  F_{ph}^{(\nu)} = \sum_n X_{ph}^{(n)} F_n^{(\nu)}.
\end{equation}
Similar transformation can be done for the $Y$ amplitudes, but as their values are very small they will not be shown here.
From this figure it can be seen that the original RPA state $\omega_n = 21.3$ MeV is a very collective state with many 1p-1h components involved in.
When considering the escape and spreading effect of the PVC, we noticed that this state becomes even more collective.
In both PVC and PVC-dia calculations, the wave functions $F_{ph}^{(\nu)}$ of these states are similar.

\begin{figure}[htbp]
  \includegraphics[width=8cm]{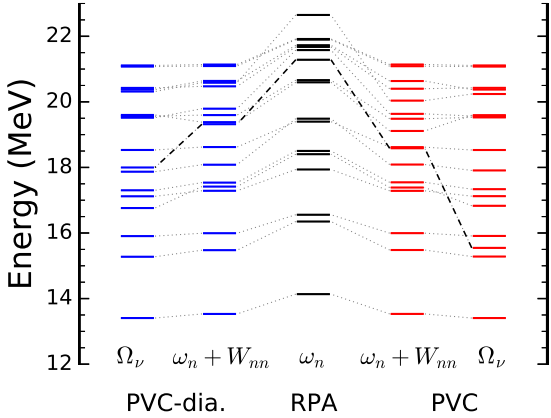}
  \caption{(Color online) Excitation energies of ISGQR in $^{16}$O calculated with SAMi functional by RPA, PVC, and PVC-dia.
  The diagonal PVC matrix elements before diagonalizing the PVC Hamiltonian $\mathcal{H}_{nn}$ in Eq.~(\ref{eq:hnn}) are also given.}
  \label{fig:wn}
\end{figure}

After identifying the major RPA components of the PVC states, one can see how the eigenenergies change from RPA to PVC.
In Fig.~\ref{fig:wn} we show the RPA solutions $\omega_n$ for the ISGQR of $^{16}$O, 
below 23 MeV, obtained by using the SAMi functional, together with the corresponding eigenenergies of 
the PVC solutions at $\omega = 14.6$ MeV.
The diagonal matrix elements of the PVC Hamiltonian (\ref{eq:hnn}) before diagonalizing, $\mathcal{H}_{nn}(\omega) = \omega_n + W_{nn}(\omega)$ have also been shown, with $W = W^\uparrow + W^\downarrow$ the escape term plus spreading term.
The corresponding levels are connected with dotted lines, with bold dashed lines emphasising the link between RPA state $\omega_n = 21.3$ MeV, and PVC states $\Omega_\nu({\rm PVC}) = 15.5$ MeV and $\Omega_\nu(\text{PVC-dia}) = 18.0$ MeV.

It can be seen from Fig.~\ref{fig:wn} that the diagonal PVC matrix elements are attractive.
In PVC-dia $W_{nn} = -2.0$ MeV while for PVC the value is $-2.7$ MeV, that is, $0.7$ MeV more attraction by removing the diagonal approximation.
After diagonalizing the PVC Hamiltonian $\mathcal{H}$, the energy level changes from the 
perturbative approximation $\mathcal{H}_{nn}$ (originated from the $\omega_n = 21.3$ MeV RPA state) to the final eigenvalue $\Omega_\nu$ with 
a further decrease of 
$1.3$ MeV in PVC-dia, and of $3.1$ MeV in PVC.
In the end, the eigenenergy of this state in PVC-dia is $\Omega_\nu(\text{PVC-dia}) = 21.3 - 2.0 - 1.3 = 18.0$ MeV, while in PVC is $\Omega_\nu({\rm PVC}) = 21.3 - 2.7 - 3.1 = 15.5$ MeV.

\begin{figure}[htbp]
  \includegraphics[width=8cm]{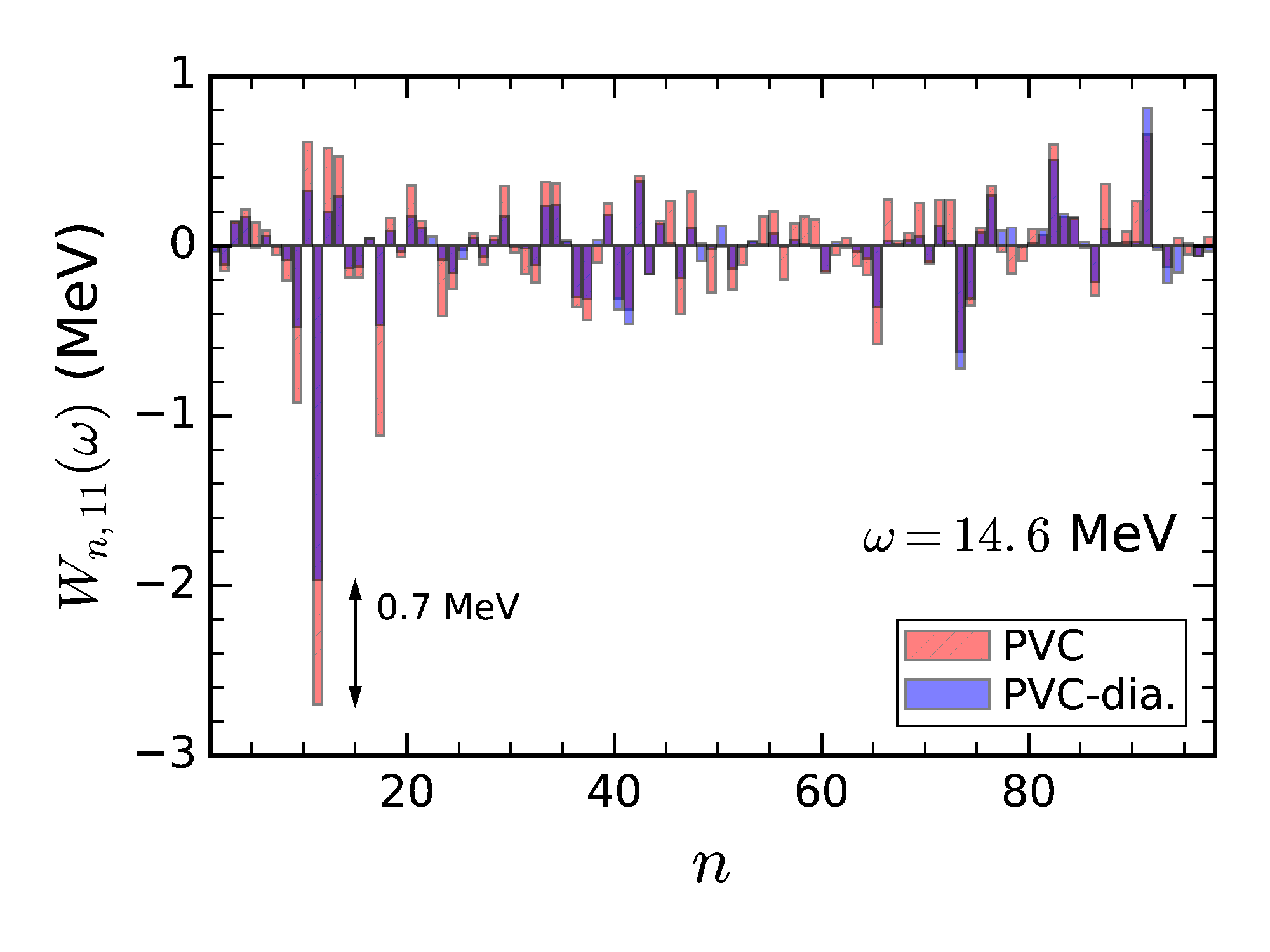}
  \caption{(Color online) PVC matrix elements $W = W^\uparrow + W^\downarrow$ with index $n$ referring to the RPA basis.
  Results are shown for PVC and PVC-dia for the ISGQR of $^{16}$O at excitation energy $\omega = 14.6$ MeV.
  The 11th RPA state is the one with excitation energy $\omega_{n=11} = 21.3$ MeV and being discussed in the text.}
  \label{fig:me}
\end{figure}

In Fig.~\ref{fig:me}, the PVC matrix elements $W$ are shown with the index $n$ referring to the RPA basis.
Since the numbering for the RPA state we are interested in is $n = 11$, with excitation energy $\omega_n = 21.3$ MeV, the matrix elements are shown for $W_{n,11}$.
In this figure, the big attraction of the diagonal matrix elements $W_{11,11}$ for both calculations can be clearly seen, with $0.7$ MeV more in PVC calculation.
Moreover, the magnitudes of the nondiagonal matrix elements are generally larger PVC calculation, which in the end leads to more mixing of other states and lower eigenvalues after diagonalizing.

As a conclusion, the extra attraction shown by removing the diagonal approximation in the PVC model is the main cause of the appearance of the low energy peak in the ISGQR.

\subsection{Influence of diagonal approximation: coupling between neutron and proton particle-hole configurations}\label{sec:dia2}

In this subsection, we analyze another important difference between PVC with and without diagonal approximation shown in Fig.~\ref{fig:str}, that is, the coupling of neutron 1p-1h excitations and proton 1p-1h excitations.
The low energy peak at $\omega = 16.4$ MeV in the ISGMR will be used as an example.
The integration of the strength around this energy ($16.4 \pm 0.4$ MeV) by PVC is $\int S_{\rm PVC} d\omega = 1.5$ fm$^4$, and by PVC-dia is $\int S_{\text{PVC-dia}} d\omega = 1.3$ fm$^4$.

\begin{figure}[htbp]
  \includegraphics[width=8cm]{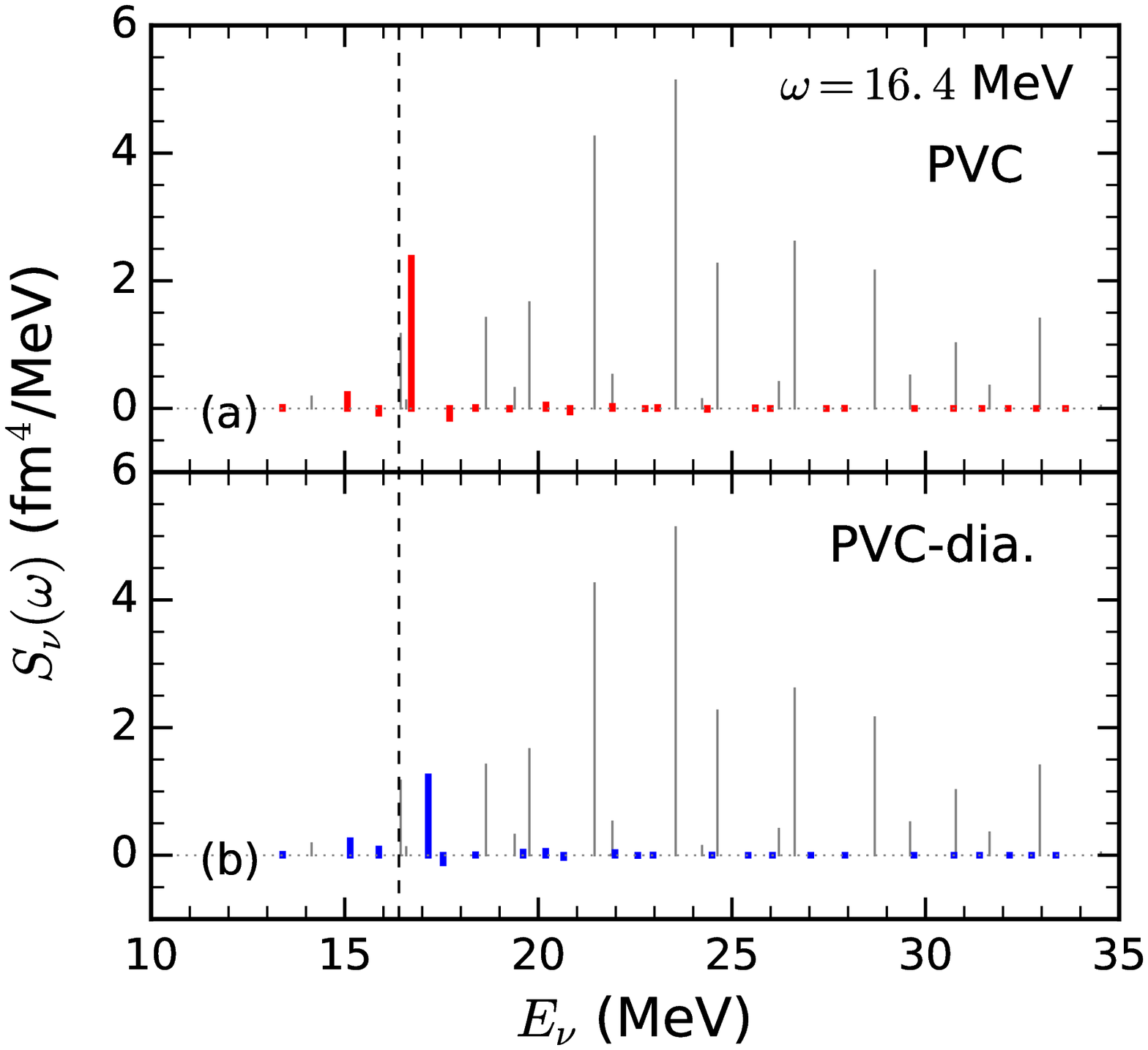}
  \caption{(Color online) Similar as Fig.~\ref{fig:strn}, but showing the contributions to the ISGMR strength function of $^{16}$O at $\omega = 16.4$ MeV from different PVC eigenstates $\nu$ with different excitation energies.
  Results of (a) PVC (red lines) and (b) PVC-dia (blue lines) are shown.
  The position of $\omega = 16.4$ MeV is given by the vertical dashed line, and the contributions from RPA states are given by the gray vertical lines (with unit fm$^4$).}
  \label{fig:strn2}
\end{figure}

Figure \ref{fig:strn2} shows the strength contributions from different PVC states $\nu$, as given by Eq.~(\ref{eq:str}).
The position of $\omega = 16.4$ MeV has been indicated by the vertical dashed line, and the 
contributions from the RPA states are given by the gray vertical lines in the background.
For PVC, the largest contribution to the strength at $\omega = 16.4$ MeV comes from the state at 
$\Omega_\nu = 16.7$ MeV while, while for PVC-dia, the largest contribution comes from the state at 
$\Omega_\nu = 17.2$ MeV.
At variance with the situation discussed in Fig.~\ref{fig:strn}, these two PVC states are both close to the energy being evaluated ($\omega = 16.4$ MeV).
Therefore, from Eq.~(\ref{eq:str}), one can hint that the difference in the strength should come from the difference in the transition matrix element in these two calculations.

\begin{figure}[htbp]
  \includegraphics[width=8cm]{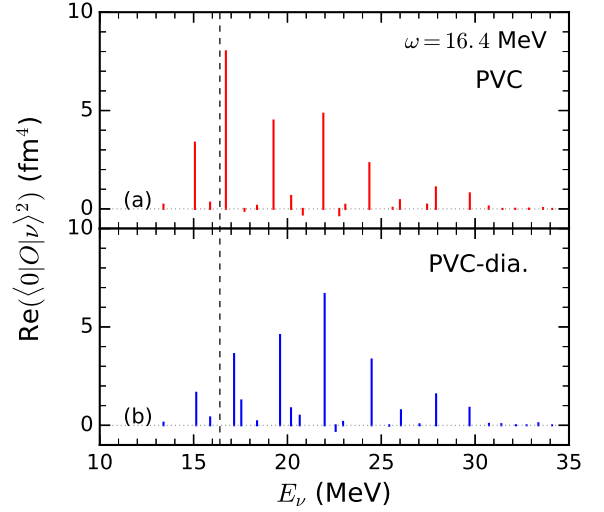}
  \caption{(Color online) Real part of the square of the transition matrix element $\langle 0|O|\nu \rangle$ for PVC states $\nu$ with different excitation energies.
  The calculation is performed for the ISGMR of $^{16}$O at $\omega = 16.7$ MeV (vertical dashed line) by (a) PVC and (b) PVC-dia.}
  \label{fig:o0n2}
\end{figure}

In Fig.~\ref{fig:o0n2} the real part of the square of the transition matrix element Re($\langle 0|O|\nu \rangle^2$) is shown.
As expected, the transition matrix element of the state $\Omega_\nu({\rm PVC}) = 16.7$ MeV is larger than the state $\Omega_\nu(\text{PVC-dia}) = 17.2$ MeV, and this explains the larger strength in PVC.

\begin{figure}[htbp]
  \includegraphics[width=8cm]{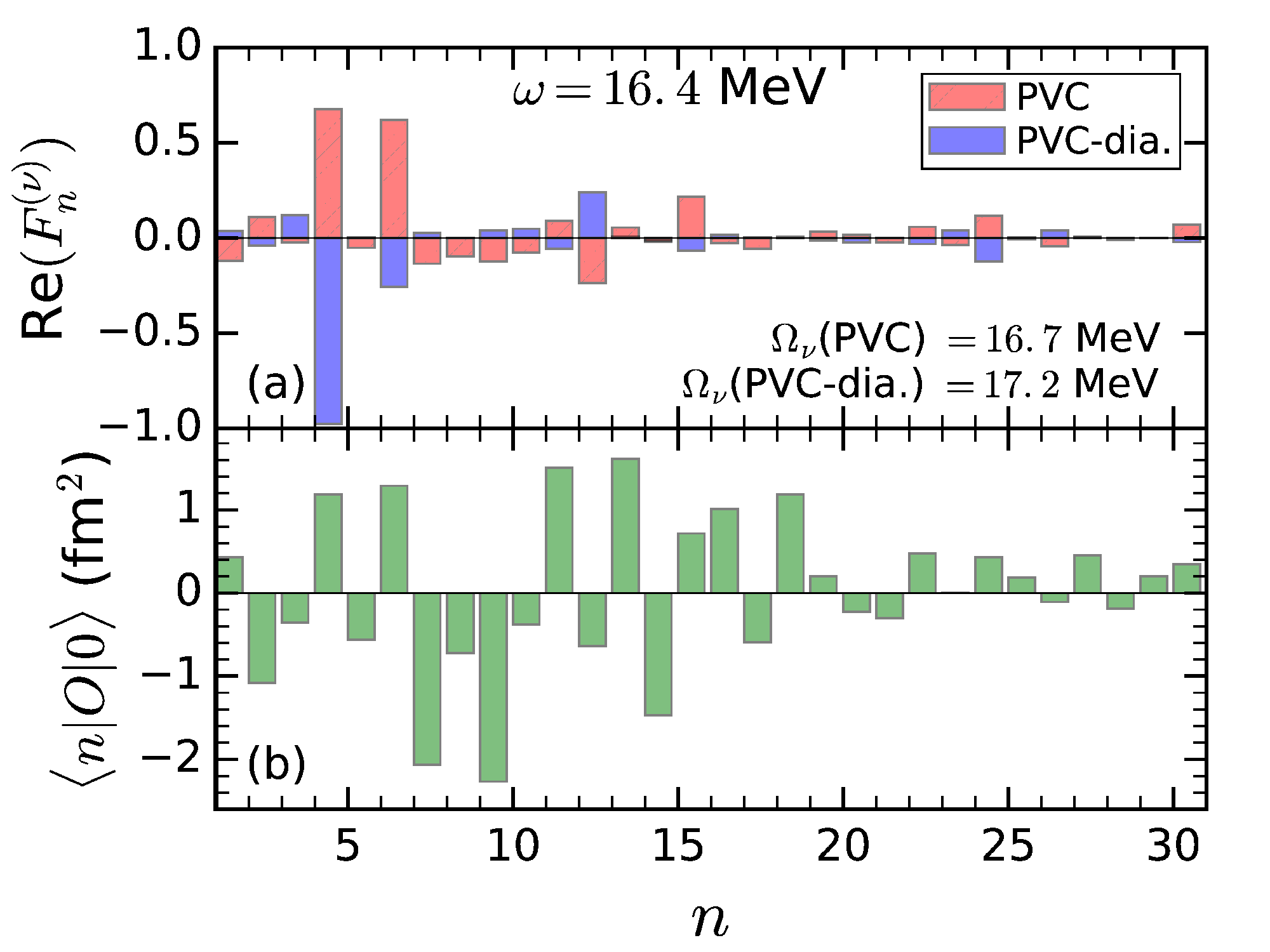}
  \caption{(Color online) (a) Real part of the PVC wave function (\ref{eq:wf}) of states $\Omega_\nu({\rm PVC}) = 16.7$ MeV and $\Omega_\nu(\text{PVC-dia}) = 17.2$ MeV in the basis of RPA states $|n\rangle$ by PVC and PVC-dia.
  (b) Transition matrix elements $\langle n|O|0 \rangle$.}
  \label{fig:wf2}
\end{figure}

To understand the difference in the transition matrix elements, we show in Fig.~\ref{fig:wf2} the PVC wave function $F_n^{(\nu)}$ (\ref{eq:wf}) of states $\Omega_\nu({\rm PVC}) = 16.7$ MeV and $\Omega_\nu(\text{PVC-dia}) = 17.2$ MeV in the basis of RPA phonons $|n\rangle$, and the transition matrix elements $\langle n|O|0 \rangle$ in the RPA representation.
The transition matrix elements $\langle 0|O|\nu \rangle$ in the PVC representation in Fig.~\ref{fig:o0n2} can be calculated 
as in Eq.~(\ref{eq:0on}).

Differently, again, from the situation of the lowest peak in the ISGQR, these two states in Fig.~\ref{fig:wf2} have very different RPA components.
In PVC-dia, the major RPA component of the state $\Omega_\nu(\text{PVC-dia}) = 17.2$ MeV can be identified as the 4th RPA phonon, while the state $\Omega_\nu({\rm PVC}) = 16.7$ MeV has the same major component but very much mixed with the 6th RPA phonon.
Since the 6th RPA phonon has a larger transition matrix element than the 4th RPA phonon, according to Eq.~(\ref{eq:0on}), the transition matrix element for $\Omega_\nu({\rm PVC}) = 16.7$ MeV is also larger.

\begin{figure}[htbp]
  \includegraphics[width=8cm]{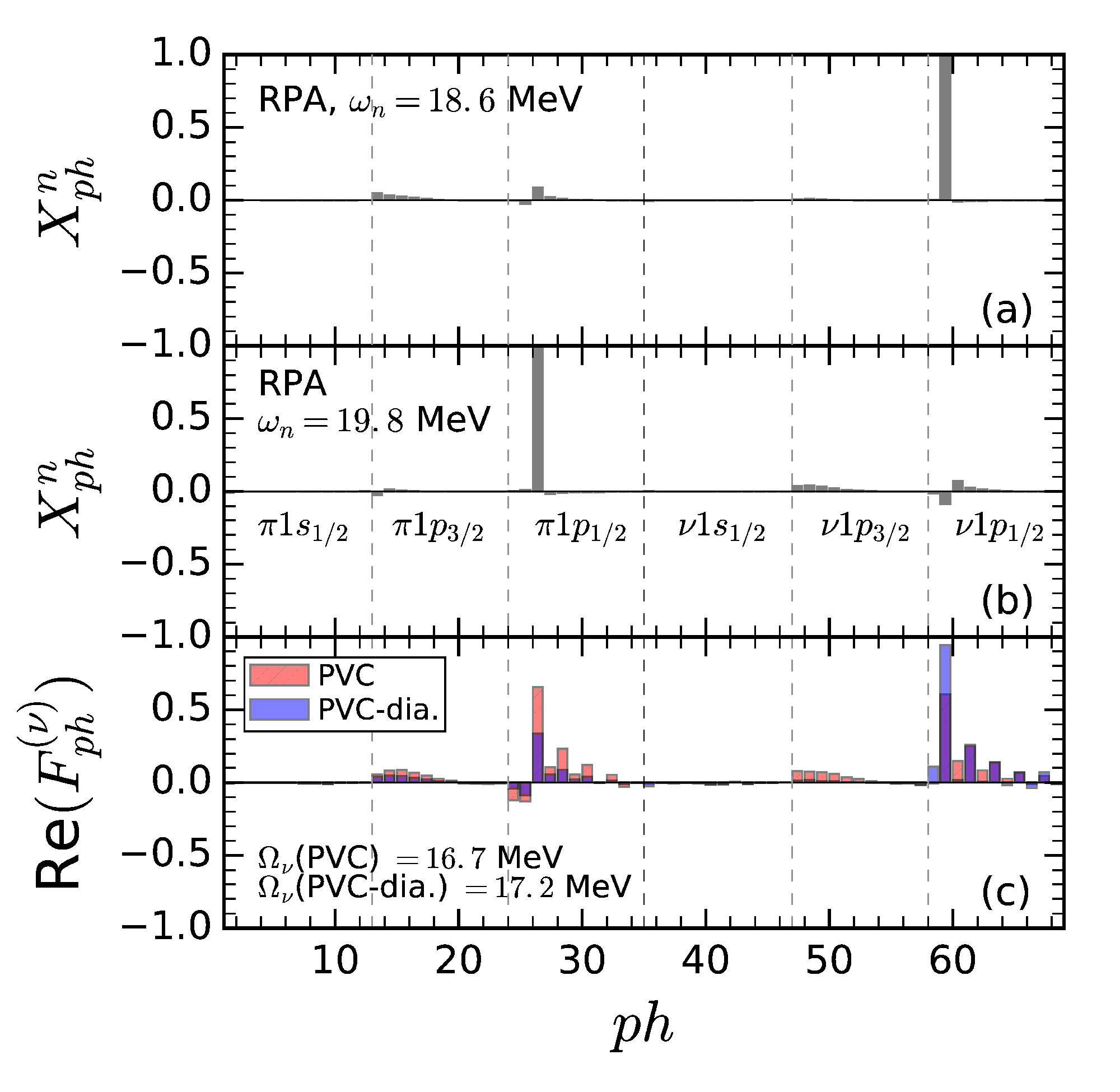}
  \caption{(Color online) RPA wave function ($X$ amplitudes) of the state with excitation energy (a) $\omega_n = 18.6$ MeV and (b) $\omega_n = 19.8$ MeV in the 1p-1h representation, see Eq.~(\ref{eq:qn}).
  Different regions divided by vertical lines are for different hole states.
  (c) Real part of the PVC wave functions of states with excitation energies $\Omega_\nu({\rm PVC}) = 16.7$ MeV and $\Omega_\nu(\text{PVC-dia}) = 17.2$ MeV, see Eq.~(\ref{eq:fph}), by PVC and PVC-dia.}
  \label{fig:wfph2}
\end{figure}

In Fig.~\ref{fig:wfph2} we show the 1p-1h components of these two PVC states as well as the related RPA states $\omega_{n=4} = 18.6$ MeV and $\omega_{n=6} = 19.8$ MeV.
As mentioned above, the PVC state $\Omega_\nu(\text{PVC-dia}) =17.2$ MeV is dominated by the RPA phonon $\omega_{n} = 18.6$ MeV and therefore its wave function in the 1p-1h representation is very similar to this phonon.
For PVC state $\Omega_\nu({\rm PVC}) = 16.7$ MeV, the RPA component $\omega_{n} = 18.6$ MeV, which is mainly a neutron $p_{1/2}$ excitation, is very much mixed with the component $\omega_{n} = 19.8$ MeV, which is mainly a proton $p_{1/2}$ excitation. 
In other words, in PVC there is a coupling between a neutron 1p-1h excitation and a proton 1p-1h excitation, which can not show up in PVC-dia as we discuss in detail in what follows.

\begin{figure}[htbp]
  \includegraphics[width=8cm]{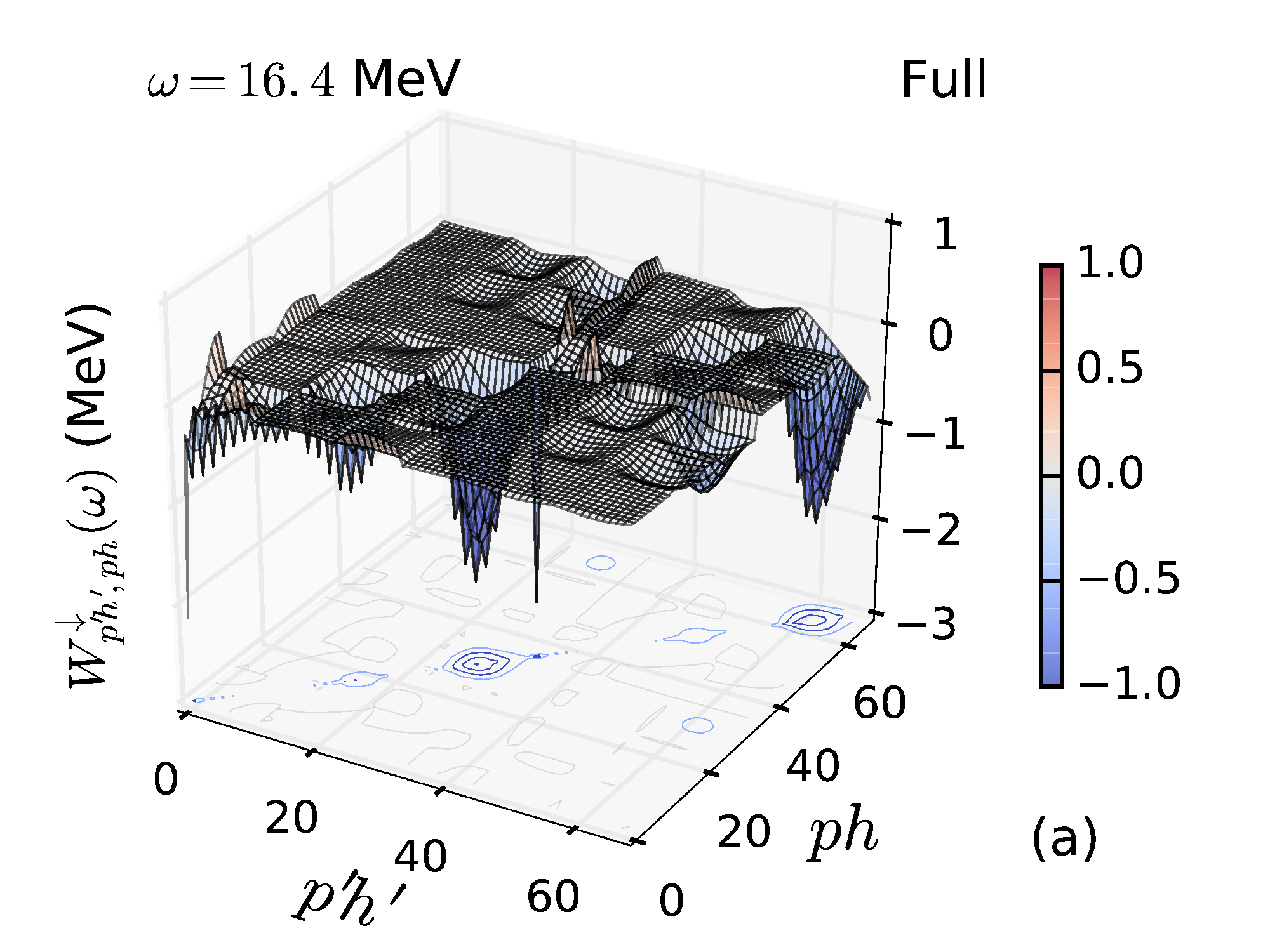}
  \includegraphics[width=8cm]{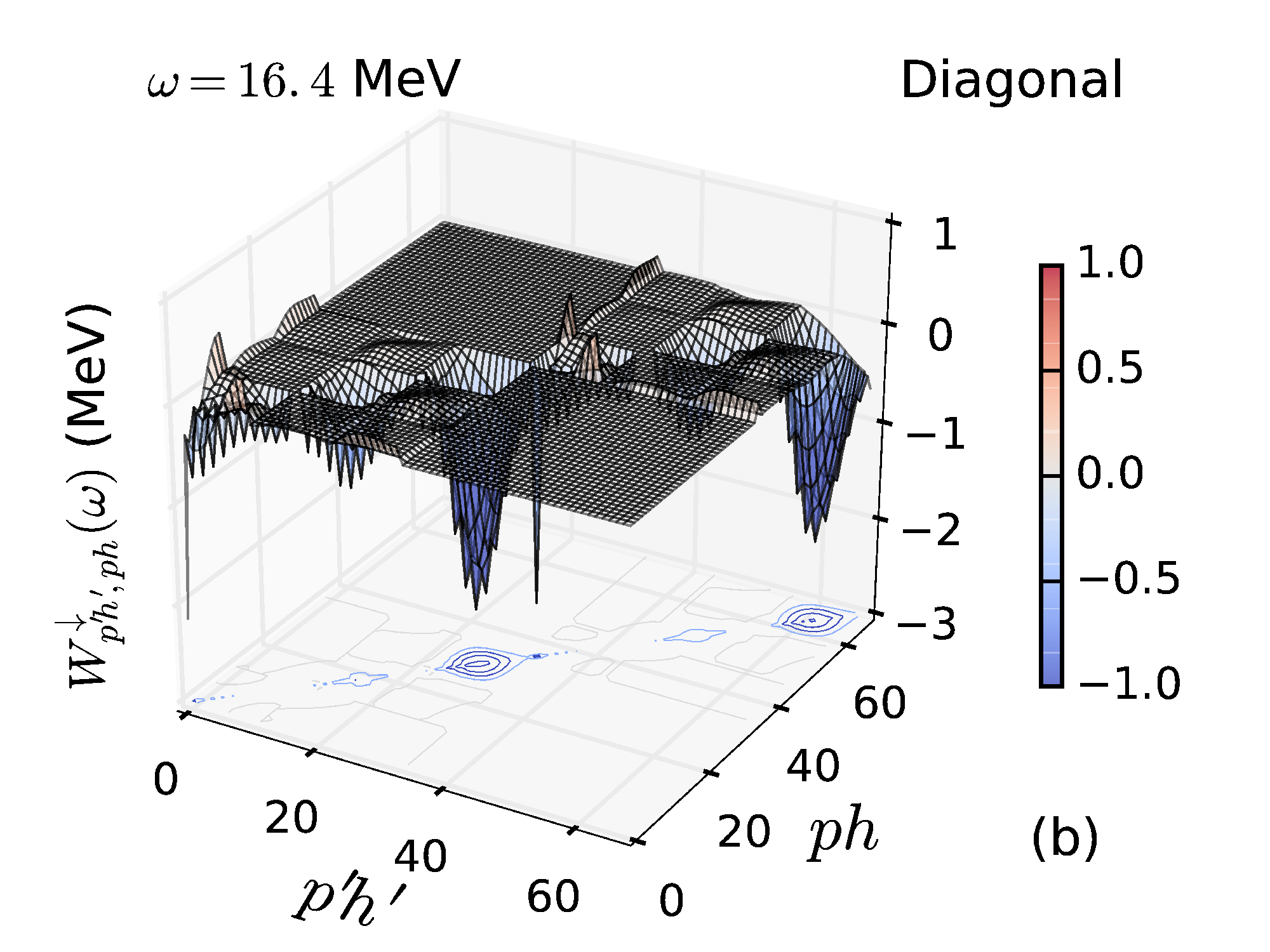}
  \caption{(Color online) Matrix elements of spreading term in the 1p-1h representation in (a) PVC (\ref{eq:wph}) and (b) PVC-dia (\ref{eq:wph2}).}
  \label{fig:wdph3d}
\end{figure}

The reason for the coupling between neutron and proton 1p-1h states is shown in Fig.~\ref{fig:wdph3d}, in which the matrix elements of the spreading term in PVC and PVC-dia are plotted.
The index $ph$ is the same as in Fig.~\ref{fig:wfph2}.
As expected, most of the matrix elements are attractive (negative values) and therefore the strength distributions are shifted to a lower energy.

The general pattern of the matrix is similar for both PVC and PVC-dia.
However, there is a clear difference that the matrix elements of neutron-proton interaction in PVC are nonzero while in PVC-dia they are zero.
This can be understood from the expression of spreading term in Eq.~(\ref{eq:wph}) and the diagram in Fig.~\ref{fig:q1hq2}.
The matrix element $A_{p'h',p_1'h_1'n}$ in Eq.~(\ref{eq:wph}) (or $Q_1HQ_2$ in Fig.~\ref{fig:q1hq2}) can not couple the initial 1p-1h excitation ($p_1'h_1'$) with the final 1p-1h excitation ($p'h'$) that has a different charge, and the same is true for $A_{p_1h_1n,ph}$.
Only in the denominator $A_{p_1'h_1'n,p_1h_1n}$ (or $Q_2HQ_2$ in Fig.~\ref{fig:q1hq2}) there is interaction $(\bar{V}_{p_1h_2h_1p_2})$ between the initial and final 1p-1h excitations with different charges.
When the diagonal approximation is applied in the denominator $Q_2HQ_2$, this interaction is removed and as a consequence the spreading term has zero matrix elements in the off-diagonal blocks where the neutron and proton 1p-1h excitations interact.

In the case of ISGQR  discussed in Fig.~\ref{fig:wfph}, the original RPA phonon is already composed of many neutron and proton 1p-1h excitations. Therefore in that case the coupling between 1p-1h states of different charges nature via the denominator in Eq.~(\ref{eq:wph}) is not significant.

\subsection{Dependence of different functionals}

\begin{figure*}[htbp!]
  \hspace{-0.5cm}
  \includegraphics[width=6.3cm]{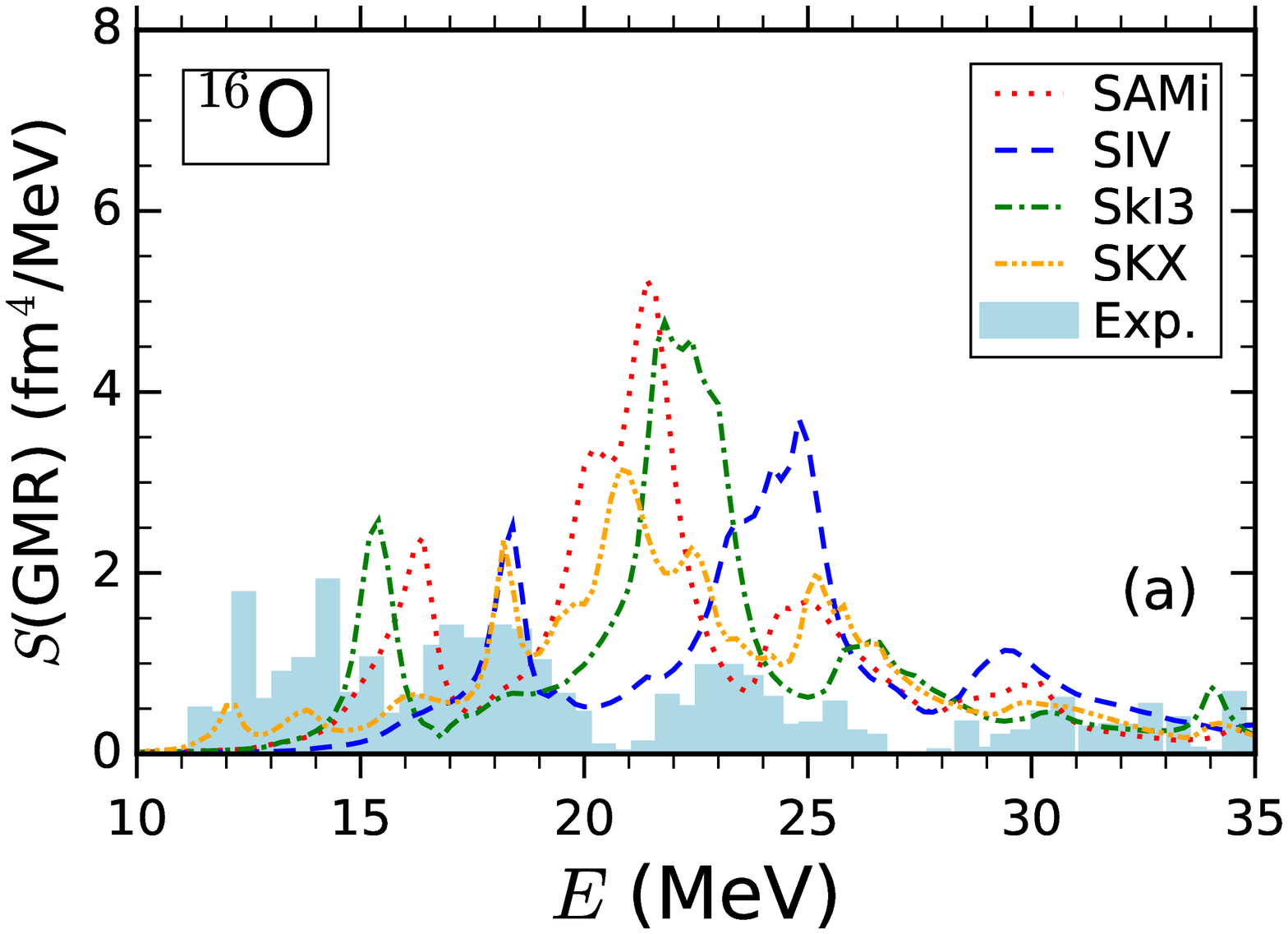}
  \hspace{-0.5cm}
  \includegraphics[width=6.3cm]{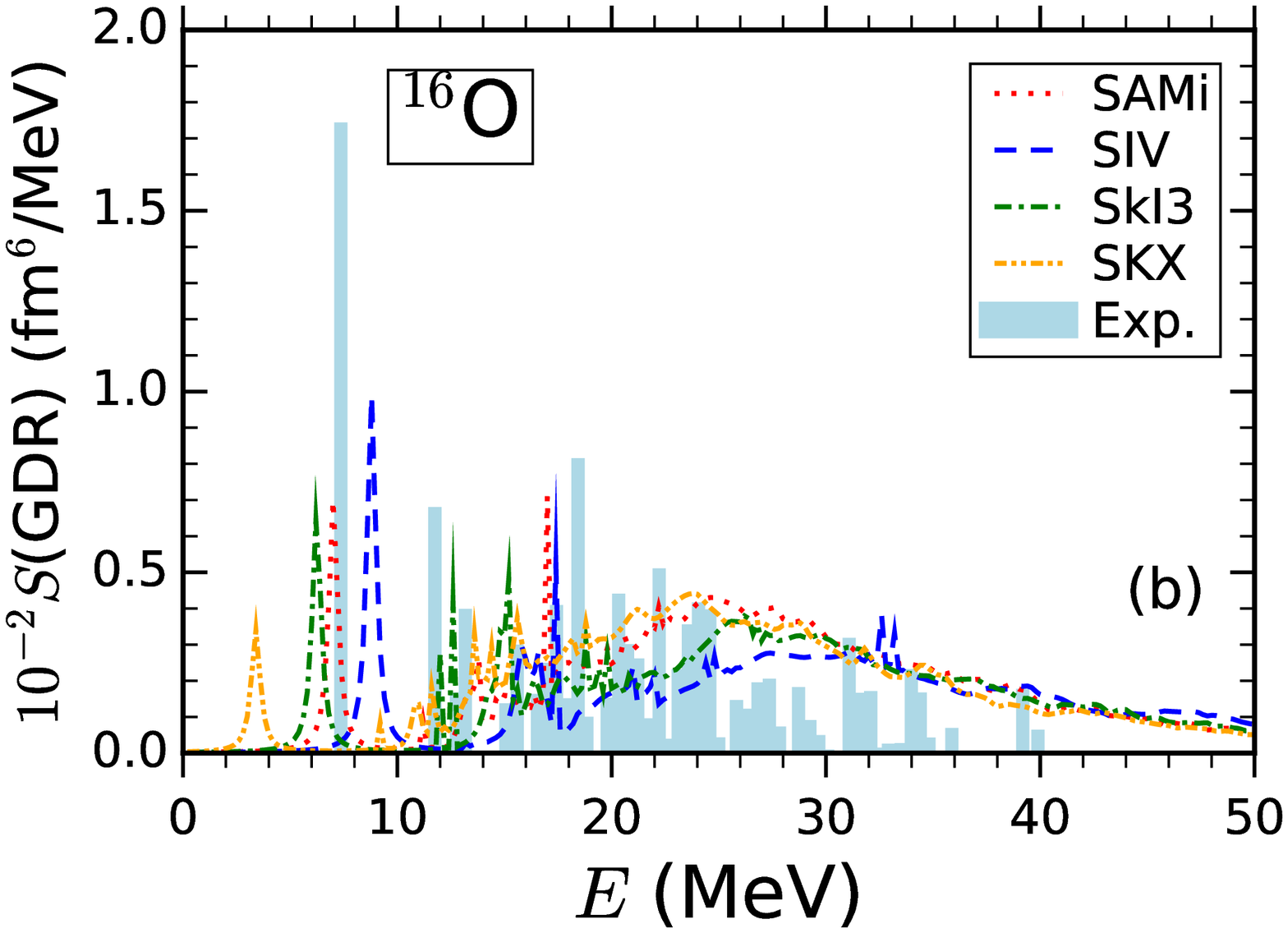}
  \hspace{-0.5cm}
  \includegraphics[width=6.3cm]{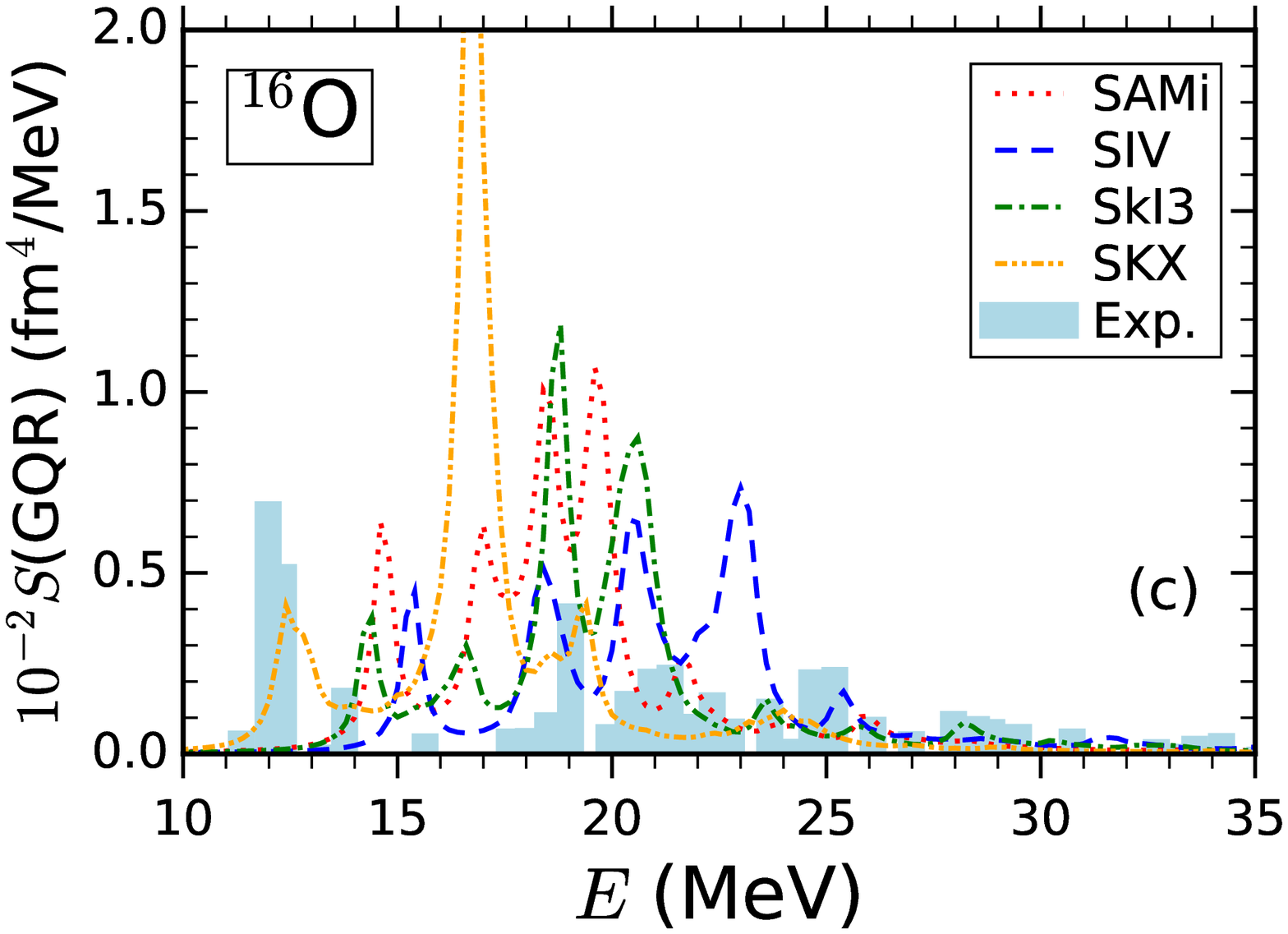} \\
  \hspace{-0.5cm}
  \includegraphics[width=6.3cm]{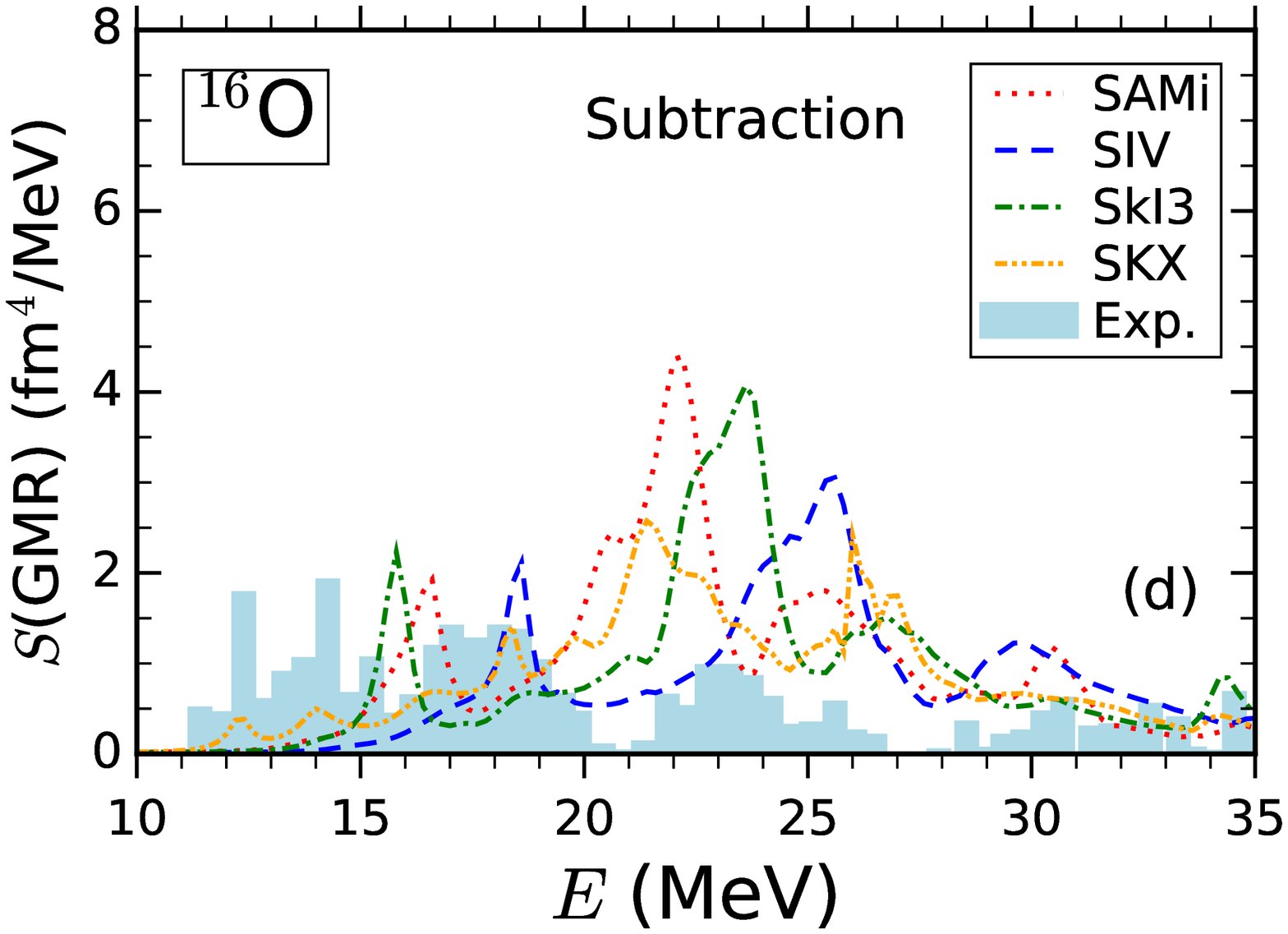}
  \hspace{-0.5cm}
  \includegraphics[width=6.3cm]{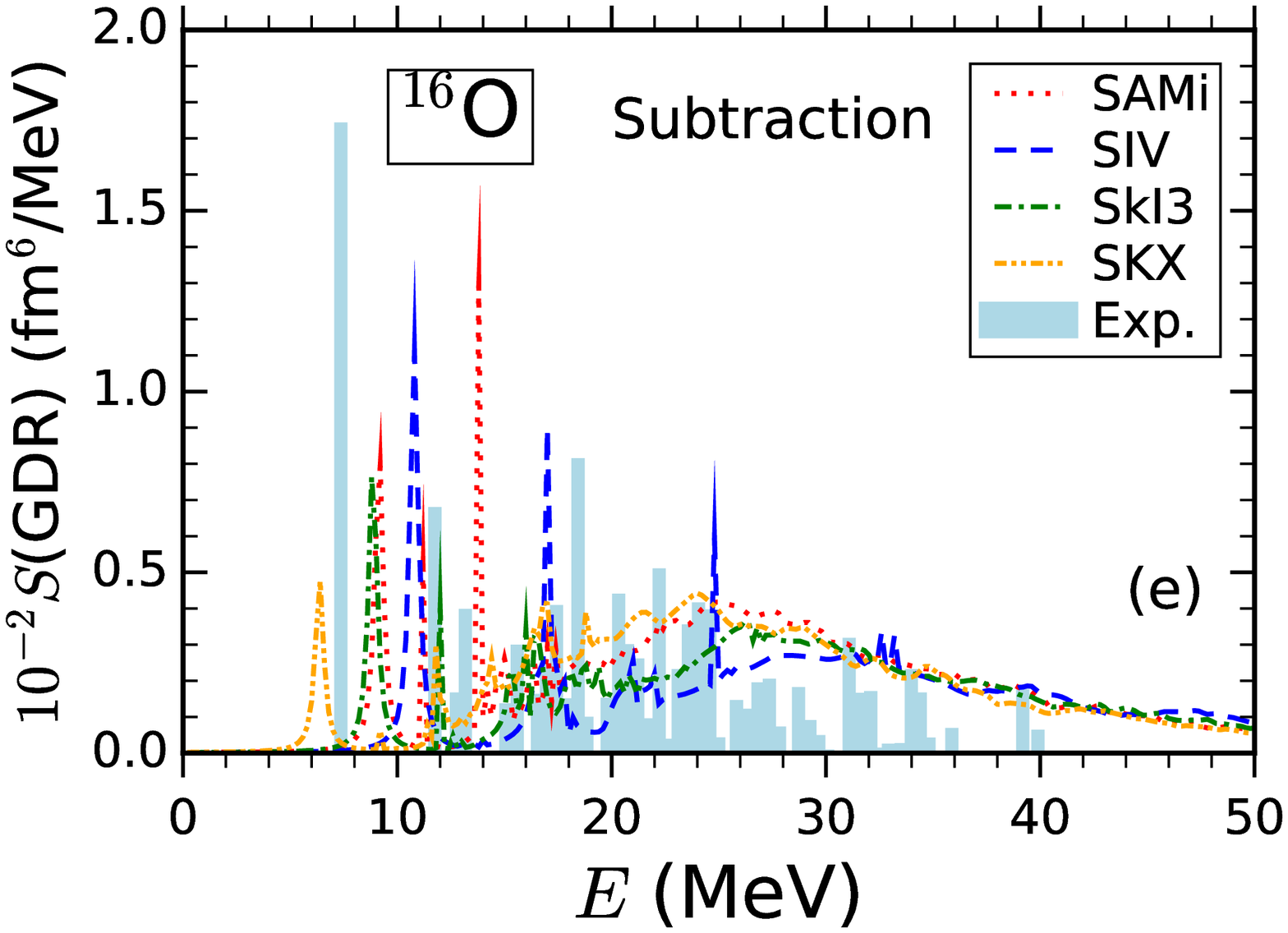}
  \hspace{-0.5cm}
  \includegraphics[width=6.3cm]{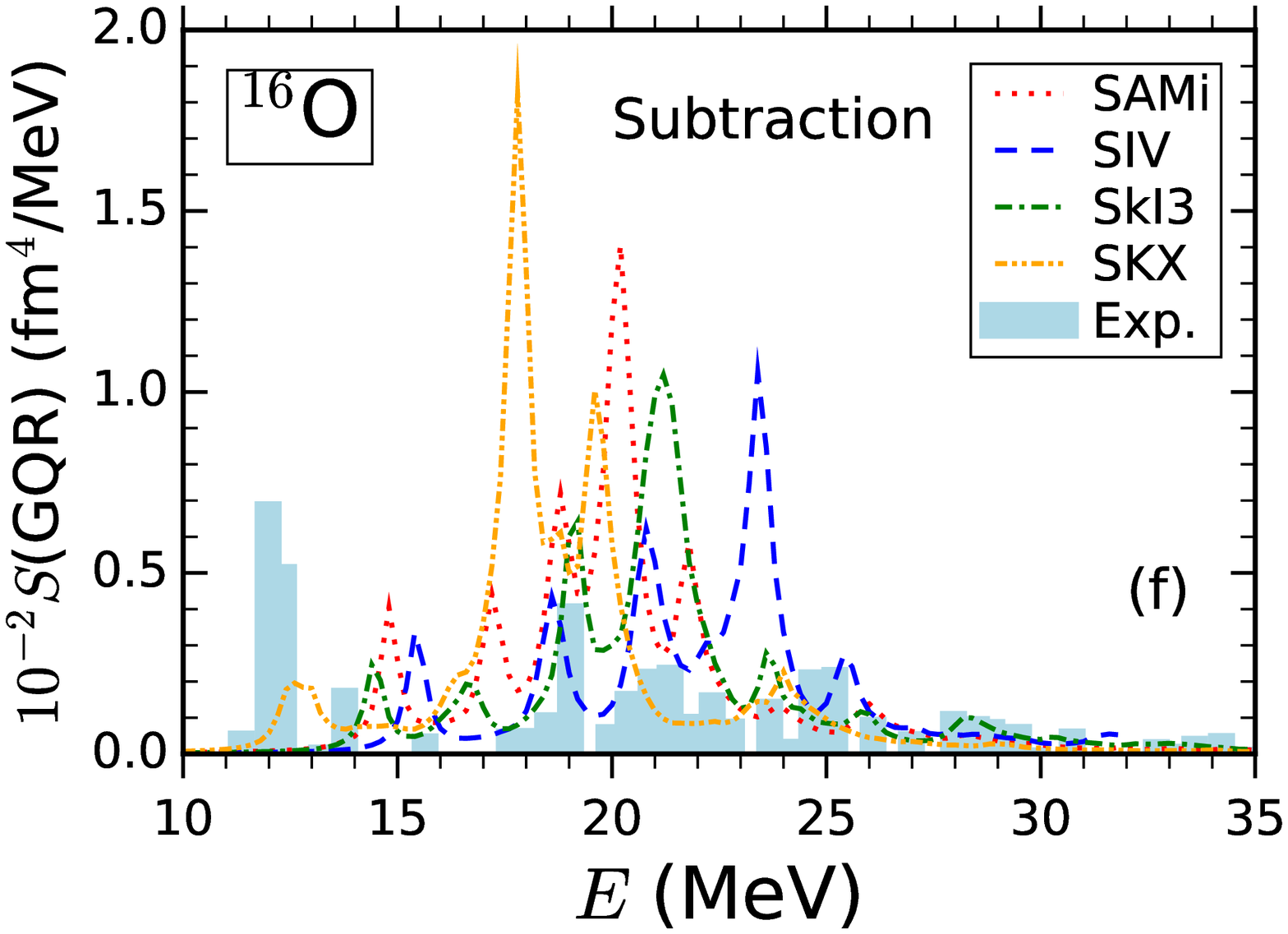}
  \caption{(Color online) Strength function of ISGMR (a,d), ISGDR (b,e), and ISGQR  (c,f) in $^{16}$O calculated by PVC without (a-c) and with (d-f) subtraction using different Skyrme functionals.
  See Fig.~\ref{fig:str} for the detail of the experimental data \cite{Harakeh1981,Lui2001}.}
  \label{fig:str3}
\end{figure*}

To test the dependence of the results on the choice of different functionals, in Fig.~\ref{fig:str3} we show the strength function of ISGMR, ISGDR, and ISGQR  in $^{16}$O calculated by PVC with and without subtraction using different Skyrme functionals: SAMi \cite{Roca-Maza2012}, SIV \cite{Beiner1975}, SkI3 \cite{Reinhard1995}, and SKX \cite{Brown1998}.
As it can be seen from the figure, the basic features such as the shape of the strength distributions obtained with different functionals are similar to each other, while in detail the results depend very much on the selected functional.
Taking the lowest peak in (a-c) of Fig.~\ref{fig:str3} as an example, SKX gives the lowest energy in all three cases: around 12 MeV in ISGMR, 3.5 MeV in ISGDR, and 12.5 MeV in ISGQR;
SIV gives the highest energy, around 18 MeV in ISGMR, 9 MeV in ISGDR, and 15.5 MeV in ISGQR;
SAMi and SkI3 sit in between with SkI3 gives slightly lower energy.

For ISGMR, if the states near 12 and 14 MeV are attributed to cluster vibrations \cite{Yamada2012}, the rest of the resonance strength around 18, 24, and 31 MeV is best described by the SIV functional.
For ISGDR, SIV gives a very strong lowest excitation near 9 MeV, in agreement with the strong experimental excitation near 7 MeV; however, the strength from 12 MeV to 24 MeV by SIV is not described as well as with the other functionals.
The dependence on the functional in the case of the ISGDR when $E > 30$ MeV is very small.
In the case of the ISGQR, SIV and SkI3 give a better description, from the excitation near 15 MeV, to 19, 21, and 25 MeV.
The strong strength near 12 MeV given by SKX is in agreement with the data, but this model gives a too large strength near 17 MeV where no experimental peak appears.

When subtraction is included in the PVC calculations shown in (d-f) of Fig.~\ref{fig:str3}, the effect for different functionals are similar to the one that has been investigated in Sec. \ref{sec:str} using SAMi.
The strength distributions are generally shifted to a higher energy by about $\approx 1-2$ MeV.
For ISGMR, the first two peaks' positions given by SIV are now slightly higher than the data, while the major peak near $\approx 22-24$ MeV given by SkI3 is in good agreement with the data.
For the ISGDR, the strength distribution given by SKX has been improved, and it describes well the experimental structures near 
7, 11, and $19-22$ MeV.
In this case the results from other functionals are not as good as they were before subtraction.
For the ISGQR, the description by SkI3 is improved with subtraction and the peaks near 19 and 21 MeV are in good agreement with the data.

The correlation between the excitation energy calculated at RPA level and nuclear matter properties has been extensively studied (cf. \cite{Roca-Maza2018a} and references therein).
For instance, the compression mode ISGMR and ISGDR are correlated with the incompressibility coefficient $K_\infty$; the ISGQR is correlated with the effective mass $m^*/m$.
Such correlations still persist for the excitation energies calculated by PVC.
For example, among the four functionals, SAMi gives the smallest incompressibility coefficient with $K_\infty^{\rm (SAMi)} = 245$ MeV while SIV gives the largest $K_\infty^{\rm (SIV)} = 325$ MeV.
Accordingly, the constrained energy $E_c = \sqrt{m_1/m_{-1}}$ given by these two functionals are $E_c^{\rm (SAMi)} = 23.7$ MeV and $E_c^{\rm (SIV)} = 26.7$ MeV for ISGMR, $E_c^{\rm (SAMi)} = 26.2$ MeV and $E_c^{\rm (SIV)} = 28.5$ MeV for ISGDR.
SKX gives the largest effective mass with $m_{\rm SKX}^*/m = 0.99$ and SIV the smallest $m_{\rm SIV}^*/m = 0.47$.
Accordingly, the centroid energy of ISGQR given by SKX is 15.0 MeV while by SIV is 21.6 MeV.
These relations are also reflected in Fig.~\ref{fig:str3}.

In short, first, for current PVC calculation it is difficult to find a functional which can give a satisfactory description for all the giant resonances studied here.
Second, for different Skyrme functionals, the correlation between the excitation energy calculated by RPA and nuclear matter properties still holds for the excitation energy calculated by PVC.

\section{Summary}\label{sec:sum}

In this work we have developed the self-consistent particle-vibration model without the diagonal approximation.
The interaction between the two particle-holes inside the doorway states has been taken into account, and it is shown that it can be derived from the equation-of-motion method similar to the one in SRPA \cite{Yannouleas1987}.
Analytical comparison of this correlation with the one in TBA \cite{Tselyaev2007,Litvinova2010,Litvinova2013} is also given.
The framework has been used to study the isoscalar giant monopole, dipole, and quadrupole resonances of $^{16}$O using Skyrme functionals.
The results are compared with the second RPA \cite{Gambacurta2010,Gambacurta2015} and (relativistic) time-blocking approximation \cite{Lyutorovich2015,Litvinova2010,Litvinova2013}.

The importance of including self-consistently the full interaction in the PVC vertex has been shown, by considering the strength distributions and sum rules.
Among the different terms of the Skyrme interaction other than the central term, the spin-orbit term, which has been ignored in most previous PVC studies, plays a significant role in our current study, especially in the case of the ISGQR.

The diagonal approximation has also much influence on the strength distribution of the ISGQR in $^{16}$O.
Without diagonal approximation, the strength distribution of the ISGQR is more fragmented and wider, in better agreement with the experimental data.
A new peak near $E = 15$ MeV appears in the PVC calculation without diagonal approximation (also present at lower energies in the experimental data). Such peak has been used as an example to show the difference induced by the diagonal approximation for the eigenenergies.
For the case of ISGMR and ISGDR, the strength distributions in $^{16}$O are less influenced by the diagonal approximation; and in 
all cases, the sum rules are not influenced by the diagonal approximation.

Another important drawback of the diagonal approximation is that one implicitly neglects the possibility of coupling between neutron and proton 1p-1h excitations included in the doorway states, that is instead recovered in the PVC calculation without diagonal approximation.
This difference is more prominent in situations where two phonons are dominated, respectively, by either a neutron or a proton 1p-1h excitation, as the interaction between 1p-1h excitations with different charge in the diagonal approximation is set to zero.
When the phonon is already composed with mixed neutron and proton 1p-1h excitations, removing the diagonal approximation may not be significant.

The subtraction method, which has been developed to renormalize the effective interaction beyond RPA, has also been investigated within the framework of PVC calculations without diagonal approximation.
It solves, to some extent, the problem that the centroid of strength distributions is slightly too low compared with experimental data.

As can be seen from the formulas of the full spreading term in Eqs.~(\ref{eq:wphj}), present PVC calculation without diagonal approximation is very time consuming, especially for heavy nuclei where more ph configurations are to be considered.
In the future we will make the calculation parallelized and apply it to study heavier systems.

Although we have shown that removing the diagonal approximation is a step to be done, there is still room to improve the PVC models.
We plan to perform further investigation on the proper treatment of phonons in the doorway states.
A recent investigation within the time blocking approximation \cite{Tselyaev2018} might provide some interesting insight in this respect, as the authors propose a way to choose the most relevant phonons 
and achieve convergence with respect to the model space.
The works of \cite{Litvinova2010,Litvinova2013,Litvinova2018a} are also of particular interest as a guidance for future development.
One can first take into account the ground-state correlation and include the RPA phonon coupling among the doorway states (see discussion in Appendix \ref{app:tba}).
Further developments to compare with the equation-of-motion diagrams in \cite{Litvinova2018a} and to go beyond 2p-2h level can also be made.
At the same time, the diagonal approximation should be tested in more nuclei and, even more importantly, in the case of other types of resonances such as spin-isospin resonances.
This may impact on the problem of the Gamow-Teller quenching or on the $\beta$-decay processes
of astrophysical interest.

\section*{ACKNOWLEDGMENTS}

This work was partly supported by Funding from the European Union's Horizon 2020 research and innovation programme under Grant agreement No. 654002.

\appendix

\section{Effective Hamiltonian in $Q_1$ subspace}\label{app:hq1}

The subspaces $P,Q_1$ and $Q_2$ have the following properties
\begin{align}
  &P^2 = P, \quad Q_1^2 = Q_1, \quad Q_2^2 = Q_2, \notag \\
  &PQ_1 = PQ_2 = Q_1Q_2 = 0, \notag \\
  & P+Q_1+Q_2 = 1.
\label{eq:pqq}
\end{align}

The eigenequation
\begin{equation}\label{eq:}
  H\Psi = \omega \Psi
\end{equation}
becomes
\begin{equation}\label{eq:hpwp}
  H(P+Q_1+Q_2)\Psi = \omega (P+Q_1+Q_2)\Psi.
\end{equation}

Multiply operator $P,Q_1,Q_2$ to both side of Eq.~(\ref{eq:hpwp}) and using the properties of (\ref{eq:pqq}), one can obtain a set of equations
\begin{subequations}\label{eq:}\begin{align}
  \left( \omega - H_{PP} \right) P \Psi &= H_{PQ_1} Q_1\Psi + H_{PQ_2} Q_2\Psi, \label{eq:hpp} \\
  \left( \omega - H_{Q_1Q_1} \right) Q_1 \Psi &= H_{Q_1P} P\Psi + H_{Q_1Q_2} Q_2\Psi, \label{eq:hq11} \\
  \left( \omega - H_{Q_2Q_2} \right) Q_2 \Psi &= H_{Q_2P} P\Psi + H_{Q_2Q_1} Q_1\Psi ,\label{eq:hq22}
\end{align}\end{subequations}
with the subscript of the Hamiltonian represents, e.g., $H_{PQ_1} = PHQ_1$.
From Eqs.~(\ref{eq:hpp},\ref{eq:hq22}), one has
\begin{subequations}\label{eq:}\begin{align}
  P\Psi &= \frac{1}{\omega-H_{PP}} H_{PQ_1}Q_1\Psi + \frac{1}{\omega-H_{PP}} H_{PQ_2}Q_2\Psi, \\
  Q_2\Psi &= \frac{1}{\omega-H_{Q_2Q_2}} H_{Q_2P}P\Psi + \frac{1}{\omega-H_{Q_2Q_2}} H_{Q_2Q_1}Q_1\Psi.
\end{align}\end{subequations}
A small quantity $i\epsilon$ should be added in the denominator but has not been written out explicitly.
Substitute back into Eq.~(\ref{eq:hq11}), one obtains \cite{Roca-Maza2017}
\begin{align}
  \left( \omega - H_{Q_1Q_1} \right) Q_1 \Psi &= \left[ W^\uparrow(\omega) + W^\downarrow(\omega) + ... \right] Q_1\Psi,
\label{eq:}
\end{align}
with the expression of $W^\uparrow(\omega)$ and $W^\uparrow(\omega)$ have been given in Eq.~(\ref{eq:homega}).
Truncating the expansion to the leading order, one has Eqs.~(\ref{eq:pvceq}) and (\ref{eq:homega}).

\section{Comparing with RTBA}\label{app:tba}

In this part we compare our formalism without diagonal approximation to the formalism of phonon correlation in the RTBA \cite{Litvinova2010,Litvinova2013}, in which the one with diagonal approximation is labeled as RQTBA (Q for quasiparticle) and the one with phonon correlation as RQTBA2.
The spreading term $W^\downarrow$ (\ref{eq:wd}) in current framework corresponds to the so-called dynamic part of the interaction amplitude $\Phi$ in RQTBA \cite{Litvinova2010,Litvinova2013}, and they will be the objects we are comparing.
To be concise but without losing generality, we focus only on the first one of the spreading term $W^\downarrow$ (see Fig. \ref{fig:wdph} (e) and (a) in Appendix \ref{app:me}, for with and for without diagonal approximation).
They are to be compared with the first terms of $\Phi$ (with diagonal approximation) and $\bar{\Phi}$ (without diagonal approximation) in Fig. 3 of \cite{Litvinova2013}.
After adapting some of the notations to current paper, the dynamic part of the interaction amplitude in RQTBA and RQTBA2 are
\begin{widetext}
\begin{align}
  \Phi_{k_1k_4,k_2k_3}(1;\omega) &= \delta_{k_2k_4} \sum_{k_5n} \gamma_{n,k_1k_5} \frac{1}{\omega-\omega_n-e_{k_5k_2}} \gamma_{n,k_5k_3}, \label{eq:phi} \\
  \bar{\Phi}_{k_1k_4,k_2k_3}(1;\omega) &= \frac{1}{2} \sum_{k_5k_5',nn'} \gamma_{n,k_1k_5} \mathcal{R}_{n',k_5k_2} \frac{1}{\omega-\omega_n-\omega_{n'}} \mathcal{R}_{n',k_5'k_4}\gamma_{n,k_5k_3},
\label{eq:phi2}
\end{align}
\end{widetext}
where $\gamma_{n,k_1k_5}$ is the interaction between two quasiparticles $(k_1,k_5)$ and a phonon $(n)$;
$\mathcal{R}_{n',k_5k_2}$ is the phonon transition density.
The $1/2$ factor in the second equation due to symmetry consideration \cite{Tselyaev2007}.
The positive- and negative-frequency index of the relativistic Hartree-Bogoliubov equation in \cite{Litvinova2010,Litvinova2013} are not explicitly written out for simplicity.
The expressions for $\gamma$ and $\mathcal{R}$ are given as \cite{Litvinova2010,Litvinova2013}:
\begin{align}
  \gamma_{n,k_1k_2} &= \sum_{k_3k_4} V_{k_1k_4,k_2k_3} \mathcal{R}_{\mu,k_3k_4}, \label{eq:gamma} \\
  \mathcal{R}_{n,k_1k_2} &= \frac{1}{\omega_n-e_{k_1k_2}} \sum_{k_3k_4} V_{k_1k_4,k_2k_3} \mathcal{R}_{n,k_3k_4},
\label{eq:rxy}
\end{align}
The corresponding terms of Eqs. (\ref{eq:phi},\ref{eq:phi2}) in our framework are
\begin{widetext}
\begin{align}
  W_{p'h',ph}^{\downarrow{\rm dia}}(1;\omega) &= \delta_{h'h} \sum_{p_1n} V_{n,p'p_1} \frac{1}{\omega-\omega_n-e_{p_1h}} V_{n,p_1p}, \label{eq:wphi} \\
  W_{p'h',ph}^\downarrow(1;\omega) &= \sum_{p_1'p_1,n} V_{n,p'p_1'} \frac{1}{\omega-\omega_n-e_{p_1h}\delta_{p_1'h',p_1h}-\bar{V}_{p_1'hh'p_1}} V_{n,p_1p}.
\label{eq:wphi2}
\end{align}
\end{widetext}
The similarity between Eq. (\ref{eq:wphi}) and (\ref{eq:phi}) is easily seen, as both $\gamma$ in Eq. (\ref{eq:gamma}) and $V$ in Eq. (\ref{eq:vabn}) are the interactions between two (quasi)particles and a phonon.
In the following we will show the similarity between Eq. (\ref{eq:wphi2}) and (\ref{eq:phi2}).
Again, notice the term with the denominator in Eq. (\ref{eq:wphi2}) is the matrix element of the inverse of operators.
The denominator in Eq. (\ref{eq:wphi2}) can be expressed in the RPA matrix (\ref{eq:ab}) as
\begin{equation}\label{eq:oa}
  \omega-\omega_n-e_{p_1h}\delta_{p_1'h',p_1h}-\bar{V}_{p_1'hh'p_1} = \omega-\omega_n-A_{p_1'h',p_1h}.
\end{equation}
In our framework, the matrix elements of $B_{p_1h_1n_1,p_2h_2n_2}$ in Eq. (\ref{eq:b22}) is evaluated as zero.
When ground-state correlations are taken into account, this matrix elements can be nonzero and they are evaluated as $B_{p_1h_1n_1,p_2h_2n_2} = \delta_{n_1n_2} V_{p_1p_2h_1h_2}$.
Then Eq. (\ref{eq:oa}) is extended to including the RPA $B$ matrix as
\begin{equation}\label{eq:oa2}
  \omega-\omega_n-
  \left(\begin{array}{cc}
  A & B \\ -B^* & -A^*
  \end{array}\right)_{p_1'h',p_1h}
\end{equation}
Defining the following notation
\begin{align}
  \mathfrak{R} =
  \left(\begin{array}{cc}
  1 & 0 \\ 0 & -1
  \end{array}\right), \quad
  \mathfrak{X} =
  \left(\begin{array}{cc}
  X & Y^* \\ Y & X^*
  \end{array}\right), \notag \\
  \mathfrak{I} =
  \left(\begin{array}{cc}
  A & B \\ B^* & A^*
  \end{array}\right), \quad
  \Omega =
  \left(\begin{array}{cc}
  \omega_n & 0 \\ 0 & -\omega_n
  \end{array}\right)
\label{eq:xy}
\end{align}
Then Eq. (\ref{eq:oa2}) becomes
\begin{equation}\label{eq:oa3}
  \omega-\omega_n-\mathfrak{R}\mathfrak{I}.
\end{equation}
The RPA equation can be written as
\begin{equation}\label{eq:}
  \mathfrak{R}\mathfrak{I}\mathfrak{X} = \mathfrak{X}\Omega.
\end{equation}
The orthonormalization condition is
\begin{equation}\label{eq:}
  \mathfrak{X}^\dagger \mathfrak{R}\mathfrak{X} = \mathfrak{R}, \quad
  \mathfrak{X}\mathfrak{R}\mathfrak{X}^\dagger  = \mathfrak{R}.
\end{equation}
One can then derive
\begin{widetext}
\begin{align}
  (\omega-\omega_n-\mathfrak{R}\mathfrak{I})^{-1} &= \mathfrak{X}_{n_1}\mathfrak{R}\mathfrak{X}_{n_1}^\dagger\mathfrak{R} \left[ \mathfrak{X}_{n'}\mathfrak{R}\mathfrak{X}_{n'}^\dagger\mathfrak{R} (\omega-\omega_n) - \mathfrak{X}_{n'}\Omega_{n'}\mathfrak{R}\mathfrak{X}_{n'}^\dagger\mathfrak{R} \right]^{-1} \mathfrak{X}_{n_2}\mathfrak{R}\mathfrak{X}_{n_2}^\dagger\mathfrak{R} \notag \\
  &= \mathfrak{X}_{n_1}\mathfrak{R}\left\{ \mathfrak{X}_{n_2}^\dagger\mathfrak{R} \left[ \mathfrak{X}_{n'}\mathfrak{R}\mathfrak{X}_{n'}^\dagger\mathfrak{R} (\omega-\omega_n) - \mathfrak{X}_{n'}\Omega_{n'}\mathfrak{R}\mathfrak{X}_{n'}^\dagger\mathfrak{R} \right] \mathfrak{X}_{n_1}\mathfrak{R}\right\}^{-1} \mathfrak{X}_{n_2}^\dagger\mathfrak{R} \notag \\
  &= \mathfrak{X}_{n_1}\mathfrak{R}\left\{ \delta_{n_1n'}\delta_{n_2n'} \left[ (\omega-\omega_n) - \mathfrak{R}\Omega_{n'}\mathfrak{R} \right] \right\}^{-1} \mathfrak{X}_{n_2}^\dagger\mathfrak{R},
\label{eq:}
\end{align}
where $n_1,n_2,n'$ are summation indices.
After sum over $n_1$ and $n_2$, the above equation can be written explicitly as
\begin{equation}\label{eq:}
  (\omega-\omega_n-\mathfrak{R}\mathfrak{I})^{-1} = \sum_{n'} \mathfrak{X}_{n',p_1'h'}\mathfrak{R} (\omega-\omega_n-\omega_{n'})^{-1} \mathfrak{X}_{n',p_1h}\mathfrak{R}.
\end{equation}
Therefore the spreading term (\ref{eq:wphi2}) becomes
\begin{align}
  W_{p'h',ph}^\downarrow(1;\omega) &= \sum_{p_1'p_1,nn'} V_{n,p'p_1'} \mathfrak{X}_{n',p_1'h'}\mathfrak{R} \frac{1}{\omega-\omega_n-\omega_{n'}} \mathfrak{X}_{n',p_1h}\mathfrak{R} V_{n,p_1p}.
\label{eq:}
\end{align}
\end{widetext}
This is similar to the expression (\ref{eq:phi2}) in RQTBA2, with $\mathcal{R}$ in Eq. (\ref{eq:rxy}) and $\mathfrak{X}$ in Eq. (\ref{eq:xy}) are both phonon transition densities.
But notice in our approach, similar to SRPA, the matrix elements of $B_{p_1h_1n_1,p_2h_2n_2}$ in Eq. (\ref{eq:b22}) is evaluated as zero.
As a consequence, only the RPA $A$ matrix (or TDA matrix) appears in the denominator of Eq. (\ref{eq:wphi2}).
Therefore, similar to RQTBA2, the correlation between two 1p-1h of the doorway states is by an additional phonon coupling.
While in current framework without considering ground-state correlation, it is a TDA phonon instead of a RPA phonon in RQTBA2.

\section{Matrix elements of spreading term}\label{app:me}

Following the equation-of-motion method \cite{Rowe1968}, the PVC equation is derived in a similar way as the SRPA in Ref.~\cite{Yannouleas1987}.
With Eqs.~(\ref{eq:q1hq2},\ref{eq:q2hq2}) and Eqs.~(\ref{eq:a12}-\ref{eq:b12}), the full spreading term (\ref{eq:wd}) reads
\begin{widetext}
\begin{align}
  W_{p'h',ph}^\downarrow = \sum_{p_1'h_1'p_1h_1n}
  \left(\begin{array}{cc}
  A_{p'h',p_1'h_1'n} & 0 \\
  0 & -A_{p'h',p_1'h_1'n}^* \\
  \end{array}\right)
  \left(\begin{array}{cc}
  \frac{1}{\omega-A_{p_1'h_1'n,p_1h_1n}+i\epsilon} & 0 \\
  0 & \frac{1}{\omega+A_{p_1'h_1'n,p_1h_1n}^*+i\epsilon} \\
  \end{array}\right)
  \left(\begin{array}{cc}
  A_{p_1h_1n,ph} & 0 \\
  0 & -A_{p_1h_1n,ph}^* \\
  \end{array}\right)
\label{eq:}
\end{align}
This is a two-by-two matrix with dimension corresponding to the RPA matrix in Eq.~(\ref{eq:rpa}).
Without causing confusion we can write it as
\begin{equation}\label{eq:}
  W^\downarrow \to 
  \left(\begin{array}{cc}
  W_{p'h',ph}^\downarrow(\omega) & 0 \\
  0 & -W_{p'h',ph}^{\downarrow *}(-\omega) \\
  \end{array}\right),
\end{equation}
with $W_{p'h',ph}^\downarrow(\omega)$ given in Eq.~(\ref{eq:wph}).
When the PVC equation is solved in the RPA phonon basis, one can transform this matrix to the phonon representation by
\begin{equation}\label{eq:}
  W_{n'n}^\downarrow = \sum_{p'h',ph} \left[ W_{p'h',ph}^\downarrow(\omega) X_{p'h'}^{(n')} X_{ph}^{(n)} + W_{p'h',ph}^{\downarrow *}(-\omega) Y_{p'h'}^{(n')} Y_{ph}^{(n)} \right]
\end{equation}

For spherical nuclei, the particle-hole $jj$-coupled matrix element can be used, which is defined as
\begin{equation}\label{eq:}
  \langle 12|\bar{V}|34 \rangle^J = \sum_{m_1m_2m_3m_4} (-1)^{j_3-m_3} C_{j_1m_1j_3-m_3}^{JM} (-1)^{j_2-m_2} C_{j_4m_4j_2-m_2}^{JM} \langle 12|\bar{V}|34 \rangle.
\end{equation}
The RPA operator (\ref{eq:qn}) in the coupled form is
\begin{equation}\label{eq:qnlm}
  Q_{nLM}^\dagger = \sum_{ph} \left[ X_{ph}^{nL} A_{ph}^\dagger(LM) - Y_{ph}^{nL} A_{ph}(L\overline{M}) \right],
\end{equation}
with
\begin{align}
  A_{ph}^\dagger(LM) &= \sum_{m_pm_h} (-1)^{j_h-m_h} C_{j_pm_pj_h-m_h}^{LM} a_{pm_p}^\dagger a_{hm_h}, \\
  A_{ph}(L\overline{M}) &= \sum_{m_pm_h} (-1)^{L+M+j_h-m_h} C_{j_pm_pj_h-m_h}^{L-M} a_{hm_h}^\dagger a_{pm_p}.
\label{eq:}
\end{align}
From now on without specification, the quantum number will not include the magnetic one, for example, the summation in Eq.~(\ref{eq:qnlm}) do not include $m_p$ or $m_h$.
The RPA matrix (\ref{eq:ab}) in the $jj$-coupled form (with coupled total angular momentum $J$) simply becomes
\begin{equation}\label{eq:}
  A_{p'h',ph}^J = \delta_{p'h',ph} (e_p - e_h) + \bar{V}_{p'hh'p}^J, \quad
  B_{p'h',ph}^J = \bar{V}_{p'ph'h}^J.
\end{equation}
The $jj$-coupled form of the spreading term (\ref{eq:wph}) is more complicated.
We first give the $jj$-coupled form of Eq.~(\ref{eq:a22}), with coupled total angular momentum $\lambda$,
\begin{equation}\label{eq:}
  A_{p_1h_1n_1L_1,p_2h_2n_2L_2}^\lambda = \delta_{n_1L_1,n_2L_2} \left[ \delta_{p_1h_1,p_2h_2} \left( \omega_{n_1L_1} + e_{p_1h_1} \right) + \bar{V}_{p_1h_2h_1p_2}^\lambda \right],
\end{equation}
Let the inverse of matrix $\omega-A_{p_1h_1n_1L_1,p_2h_2n_2L_2}^\lambda+i\epsilon$ be labeled as $D_{11}(\omega)$, and the inverse of matrix $-\omega-A_{p_1h_1n_1L_1,p_2h_2n_2L_2}^\lambda+i\epsilon$ be labeled as $D_{22}(\omega)$, they satisfy the follow equation (take $D_{11}$ as an example)
\begin{equation}\label{eq:inv}
  \sum_{p_1h_1} \left[ \delta_{p_1'h_1',p_1h_1} \left( \omega+i\epsilon -\omega_{nL} - e_{p_1h_1} \right) - \bar{V}_{p_1'h_1h_1'p_1}^\lambda \right] \langle p_1h_1|D_{11}(\omega)|p_2h_2 \rangle_{nL}^{\lambda} = \delta_{p_1'h_1',p_2h_2}.
\end{equation}
Since $D_{22}$ can be obtained in the same equation (\ref{eq:inv}) by simply replacing $\omega$ to $-\omega$, we will not distinguish these two matrix explicitly and simply write $\langle p_1h_1|D(\omega)|p_2h_2 \rangle_{nL}^{\lambda}$.
In the end, the full spreading term (\ref{eq:wph}) in the $jj$-coupled form can be written as
\begin{equation}\label{eq:}
  W_{p'h',ph}^{\downarrow J}(\omega) = \sum_{k = 1}^4 W_{p'h',ph}^{\downarrow J}(k;\omega),
\end{equation}
with
\begin{subequations}\label{eq:wphj}\begin{align}
  W_{p'h',ph}^{\downarrow J}(1;\omega) &=  \sum_{\substack{\lambda p_1'\\nLp_1}} F_1 \hat{\lambda}^2 \hat{L}^2
  \left\{\begin{array}{ccc} j_{p_1'} & j_{h'} & \lambda \\ J & L & j_{p'} \end{array}\right\}
  \left\{\begin{array}{ccc} j_{p_1} & j_{h} & \lambda \\ J & L & j_{p} \end{array}\right\}
  \langle p'|{V}|p_1',nL \rangle \langle p_1'h'|D(\omega)|p_1h \rangle_{nL}^\lambda \langle nL,p_1|{V}|p \rangle, \\
  W_{p'h',ph}^{\downarrow J}(2;\omega) &= \sum_{\substack{\lambda h_1'\\nLh_1}} F_2 \hat{\lambda}^2 \hat{L}^2
  \left\{\begin{array}{ccc} j_{h_1'} & j_{p'} & \lambda \\ J & L & j_{h'} \end{array}\right\}
  \left\{\begin{array}{ccc} j_{h_1} & j_{p} & \lambda \\ J & L & j_{h} \end{array}\right\}
  \langle h_1'|{V}|h',nL \rangle \langle p'h_1'|D(\omega)|ph_1 \rangle_{nL}^\lambda \langle nL,h|{V}|h_1 \rangle, \\
  W_{p'h',ph}^{\downarrow J}(3;\omega) &= \sum_{\substack{\lambda p_1'\\nLh_1}} F_3 \hat{\lambda}^2 \hat{L}^2
  \left\{\begin{array}{ccc} j_{p_1'} & j_{h'} & \lambda \\ J & L & j_{p'} \end{array}\right\}
  \left\{\begin{array}{ccc} j_{h_1} & j_{p} & \lambda \\ J & L & j_{h} \end{array}\right\}
  \langle p'|{V}|p_1',nL \rangle \langle p_1'h'|D(\omega)|ph_1 \rangle_{nL}^\lambda \langle nL,h|{V}|h_1 \rangle, \\
  W_{p'h',ph}^{\downarrow J}(4;\omega) &= \sum_{\substack{\lambda h_1'\\nLp_1}} F_4 \hat{\lambda}^2 \hat{L}^2
  \left\{\begin{array}{ccc} j_{h_1'} & j_{p'} & \lambda \\ J & L & j_{h'} \end{array}\right\}
  \left\{\begin{array}{ccc} j_{p_1} & j_{h} & \lambda \\ J & L & j_{p} \end{array}\right\}
  \langle h_1'|{V}|h',nL \rangle \langle p'h_1'|D(\omega)|p_1h \rangle_{nL}^\lambda \langle nL,p_1|{V}|p \rangle.
\end{align}\end{subequations}
Schematic diagrams for these terms are shown in Fig.~\ref{fig:wdph} (a-d), in which the meaning of the symbols are similar to those in Fig.~\ref{fig:q1hq2}.
Straight lines are for fermions (with uparrow a particle and dowarrow a hole), and wave lines are for phonons.
Solid circles are for phonon vertices $\langle i|V|j,n \rangle$ (\ref{eq:vabn}), and empty circles are for TDA phonon transition densities (see Appendix \ref{app:tba}).

\begin{figure*}[htbp]
  \includegraphics[width=8cm]{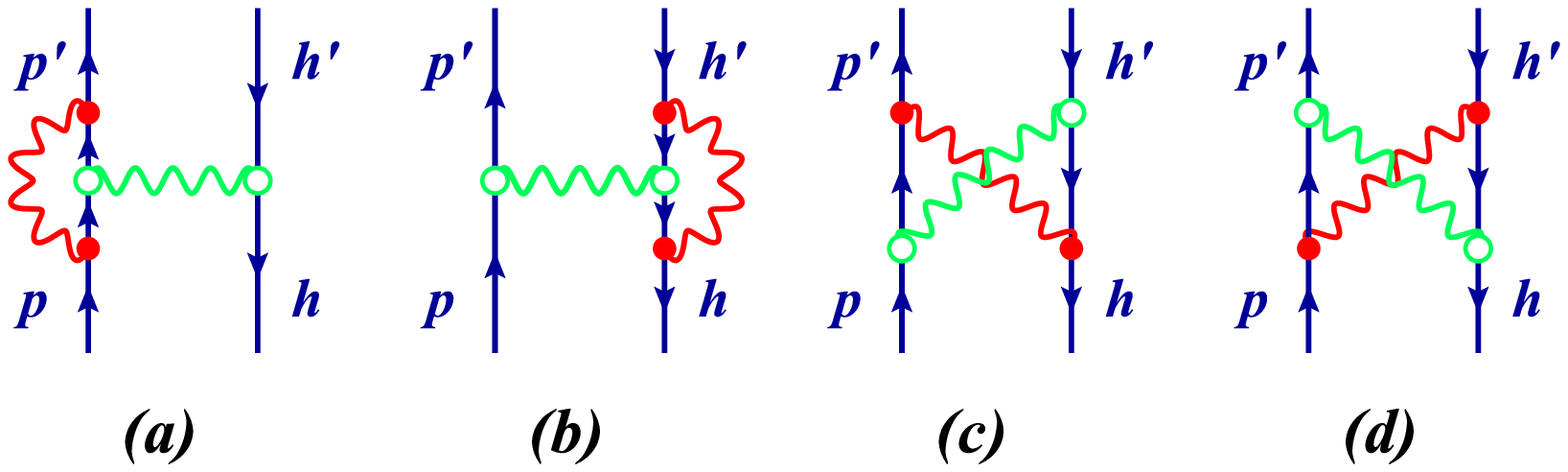}
  \hspace{1em}
  \includegraphics[width=8cm]{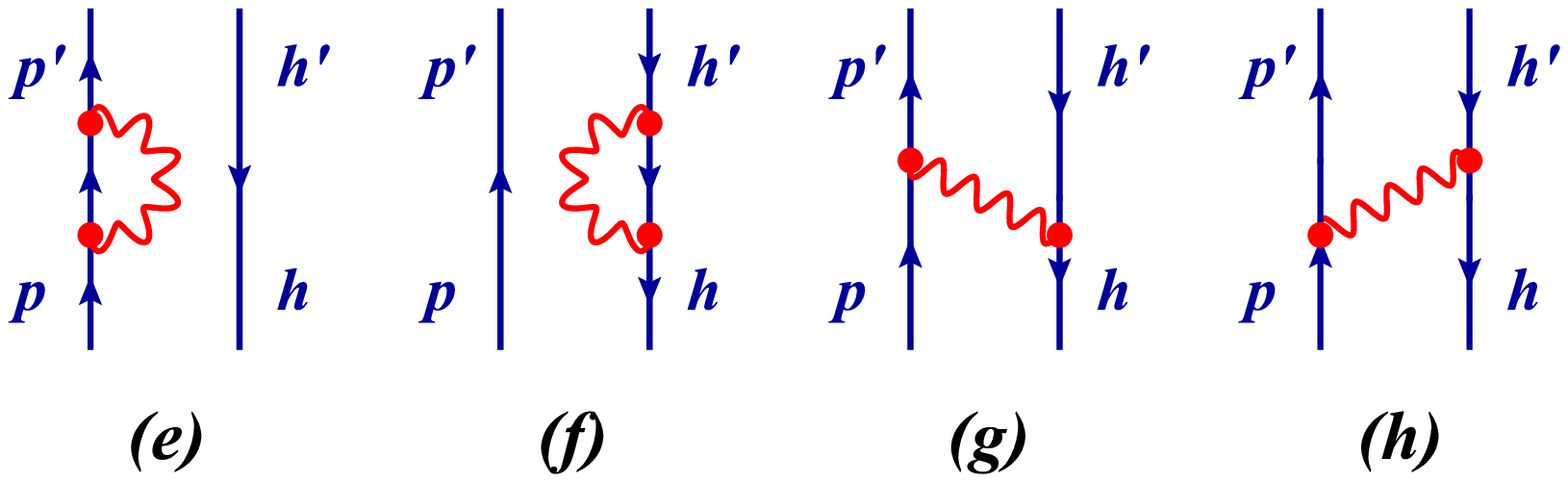}
  \caption{(a-d) Diagrammatic show of the four terms $W_{p'h',ph}^{\downarrow J}(k;\omega)$ without diagonal approximation in Eq.~(\ref{eq:wphj}); (e-h) with diagonal approximation in Eq.~(\ref{eq:wphj2}).
  Straight lines are for fermions (with uparrow a particle and dowarrow a hole), and wave lines are for phonon states. Solid circles are for phonon vertices (\ref{eq:vabn}), and empty circles are for TDA phonon transition densities (see Appendix \ref{app:tba}).}
  \label{fig:wdph}
\end{figure*}

In the above equation, $\hat{\lambda} = \sqrt{2\lambda+1}, \hat{L} = \sqrt{2L+1}$.
The coupled matrix element $\langle a|V|b,nL \rangle$ is different from the general expression in Eq.~(\ref{eq:vabn}) by
\begin{subequations}\label{eq:}\begin{align}
  \langle a|V|b,nL \rangle &= \sum_{ph} \left[ X_{ph}^{nL} \bar{V}_{ahbp}^L + (-1)^{L+j_p-j_h} Y_{ph}^{nL} \bar{V}_{apbh}^L \right], \\
  \langle nL,a|V|b \rangle &= \sum_{ph} \left[ (-1)^{L+j_p-j_h} X_{ph}^{nL} \bar{V}_{apbh}^L + Y_{ph}^{nL} \bar{V}_{ahbp}^L \right].
\end{align}\end{subequations}
The phases in the above equations are
\begin{align}
  F_1 &= (-1)^{j_{p'}+j_{h'}+j_h+L+j_{p_1}}, \quad
  &F_2 &= (-1)^{j_{p'}+j_{h'}+j_p+L+j_{h_1}} , \notag \\
  F_3 &= - (-1)^{J+j_{p'}+j_{h'}+j_p+\lambda+j_{h_1}}, \quad
  &F_4 &= - (-1)^{J+j_{p'}+j_{h'}+j_h+\lambda+j_{p_1}}.
\label{eq:}
\end{align}
When the diagonal approximation is adopted, one has
\begin{equation}\label{eq:}
  \langle p_1'h_1'|D(\omega)|p_1h_1 \rangle_{nL}^{\lambda} = \delta_{p_1'h_1',p_1h_1} \frac{1}{\omega-(\omega_{nL}+e_{p_1h_1})+i\epsilon}.
\end{equation}
There is no longer $\lambda$ dependence of matrix $D$, and the spreading terms in Eq.~(\ref{eq:wphj}) can be reduced to
\begin{subequations}\label{eq:wphj2}\begin{align}
  W_{p'h',ph}^{\downarrow J}(1;\omega) &= \delta_{h'h}\delta_{j_{p'}j_p} {\sum_{nLp_1}} (-1)^{L+j_p-j_{p_1}}\frac{\hat{L}^2}{\hat{j}_p^2} \frac{\langle p'|{V}|p_1,nL \rangle\langle nL,p_1|{V}|p  \rangle}{\omega -(\omega_{nL}+e_{p_1h}) +i\epsilon } \\
  W_{p'h',ph}^{\downarrow J}(2;\omega) &= \delta_{p'p}\delta_{j_{h'}j_h} {\sum_{nLh_1}} (-1)^{L+j_h-j_{h_1}}\frac{\hat{L}^2}{\hat{j}_h^2} \frac{\langle h_1|{V}|h',nL \rangle\langle nL,h|{V}|h_1  \rangle}{\omega -(\omega_{nL}+e_{ph_1}) +i\epsilon } \\
  W_{p'h',ph}^{\downarrow J}(3;\omega) &= -(-1)^{J+j_{p}+j_{h}} {\sum_{nL}} \hat{L}^2
  \left\{\begin{array}{ccc} j_{p} & L & j_{p'} \\ j_{h'} & J & j_h \end{array}\right\}
  \frac{\langle p'|{V}|p,nL \rangle \langle nL,h|{V}|h' \rangle}{\omega -(\omega_{nL}+  e_{ph'}) +i\epsilon } \\
  W_{p'h',ph}^{\downarrow J}(4;\omega) &= -(-1)^{J+j_{p}+j_{h}} {\sum_{nL}} \hat{L}^2
  \left\{\begin{array}{ccc} j_{p} & L & j_{p'} \\ j_{h'} & J & j_h \end{array}\right\}
  \frac{\langle h|{V}|h',nL \rangle \langle nL,p'|{V}|p \rangle}{\omega -(\omega_{nL}+ e_{p'h}) +i\epsilon } .
\end{align}\end{subequations}
They are in agreement with previous studies \cite{Colo1994}.
Schematic diagramms for these terms are shown in Fig.~\ref{fig:wdph} (e-h).
\end{widetext}


%
%

\end{CJK}
\end{document}